\documentclass[12pt,3p,authoryear]{elsarticle}

\usepackage{tikz}
\usetikzlibrary{decorations.pathreplacing}
\usepackage{hyperref}
\usepackage[utf8]{inputenc}
\usepackage{graphicx}
\usepackage{amsmath}
\usepackage{amsfonts}
\usepackage{amssymb}
\usepackage{amsthm}
\usepackage{dsfont}
\usepackage{booktabs}
\usepackage{multirow}
\usepackage{subcaption}

\usepackage[utf8]{inputenc}

\usepackage{bigstrut}

\journal{Electric Power Systems Research}
\begin{document}

\begin{frontmatter}


\title{Probabilistic forecasting with a hybrid Factor-QRA approach: Application to electricity trading}


\author[wroclaw]{Katarzyna Maciejowska}
\ead{katarzyna.maciejowska@pwr.edu.pl}
\author[wroclaw]{Tomasz Serafin \corref{cor1}}
\ead{t.serafin@pwr.edu.pl}
\author[wroclaw]{Bartosz Uniejewski}
\ead{bartosz.uniejewski@pwr.edu.pl}
\address[wroclaw]{Department of Operations Research and Business Intelligence, Wrocław University of Science and Technology, 50-370 Wrocław, Poland}
\cortext[cor1]{Corresponding author}

\begin{abstract}

This paper presents a novel hybrid approach for constricting probabilistic forecasts that combines both the Quantile Regression Averaging (QRA) method and the factor-based averaging scheme. The performance of the approach is evaluated on data sets from two European energy markets - the German EPEX SPOT and the Polish Power Exchange (TGE). The results show that the newly proposed method outperforms literature benchmarks in terms of statistical measures: the empirical coverage and the Christoffersen test for conditional coverage. Moreover, in line with recent literature trends, the economic value of forecasts is evaluated based on the trading strategy using probabilistic price predictions to optimize the operation of an energy storage system. The results suggest that apart from the use of statistical measures, there is a need for the economic evaluation of forecasts.
\end{abstract}

\end{frontmatter}


\section{Introduction}


The trading of electricity in wholesale electricity markets is nowadays a very demanding task. Market participants are exposed to various risks associated with changing weather conditions \citep{man:etal:23}, intermittent generation, and demand fluctuations that lead to an increase of the electricity price volatility \citep{ket:14, mac:20}. As a result, generators and electricity buyers, who typically need to make transaction decisions in advance, are acting under a significant hazard. Therefore, modeling and forecasting electricity markets, in particular electricity prices, are of interest to many researchers and practitioners \citep{cia:mun:zar:22,mar:nar:wer:zie:23,zha:etal:22}.

 In the literature, there are many articles that focus on point forecasts of electricity prices \citep[see][for a comprehensive review]{wer:14,pet:etal:22}. However, recent publications \citep{bil:etal:23, hon:etal:20:OAJPE, now:wer:18} and the wide interest in the 2014 Global Energy Forecasting Competition (GEFCom2014) show that more comprehensive approaches are needed, such as probabilistic forecasting. An accurate approximation of the predictive distribution of the electricity price overcomes the limitations of point forecasts and allows for better risk management (accurate estimation of the Value-at-Risk of the energy portfolio; \cite{bun:and:che:wes:16}) and serves as a tool for more complex trading strategies \citep{bun:gia:kre:18,uni:wer:21,jan:woj:22}. Various methods have been developed for probabilistic forecasting of electricity prices: quantile regression \citep{mac:20, mar:uni:wer:20}, resampling methods \citep{wan:etal:14b, kat:zie:18} and machine learning \citep{wan:etal:14c, mar:nar:wer:zie:23, mas:etal:21}. A comprehensive review of available procedures can be found in \cite{pet:etal:22} and \cite{now:wer:18}.

Quantile regression \citep[QR; ][]{koe:05} is one of the most popular methods of describing and forecasting an unknown distribution of a stochastic variable. It has been applied successfully in both macro and microeconomics \citep{koe:hal:01}. In the literature on electricity markets, it has been used to forecast electricity loads \citep{li:hu:cl:17} and spot electricity prices \citep{bun:and:che:wes:16, ser:uni:wer:19}. QR has been recently shown to serve as a link between point forecasts and probabilistic predictions. \cite{now:wer:15} proposed Quantile Regression Averaging (QRA) method, which uses a set of point predictions of electricity prices as an input to the QR, thus providing forecasts of a range of distribution quantiles.
The idea has been successfully applied to predict not only electricity prices, but also load \citep{zha:qua:sri:18} and wind generation \citep{wan:etal:19}. 
In a recent study, \cite{uni:wer:21} proposed the regularized version of quantile regression, where model regressors are selected using the Least Absolute Shrinkage and Selection Operator (LASSO). In addition, \cite{mar:uni:wer:20} introduced the Quantile Regression Machine (QRM), which uses an average of point forecasts as the only input to the QRA. In a subsequent study, \cite{ser:uni:wer:19} showed that QRM performed better than using individual predictions as separate inputs to QRA.

The idea of averaging different point forecasts via quantile regression has some limitations. The most important one is a high colinearity of input predictions. This property is a natural consequence of forecasting the same variable with different models and/or model specifications. To overcome this problem, \cite{mac:now:wer:16} interpreted the set of predictions as a panel and used a factor model to extract the included information. The paper explores a set of 32 point forecasts based on different autoregressive (ARX) models and describes it with just a few common factors, which next serve as an input to QRA. Based on the results, \cite{mac:now:wer:16} recommend using only one factor that represents the average level of predictions, which makes the method equivalent to the QRM approach.

Colinearity becomes a more pronounced problem when the number of predictions available for averaging increases, as in a recent article of \cite{mar:ser:wer:18}. The paper considers a panel of almost 700 point forecasts based on a single model estimated with different calibration windows. Combining rather than selecting an optimal window allows us to explore both local (short windows) and global (long windows) behavior of electricity prices. In the case of point forecasting, researchers typically pre-select the inputs to be used for averaging \citep{hub:mar:wer:19}. In \cite{mac:uni:ser:20}, a two-step factor-based approach is proposed for constructing point predictions, which summarizes the information included in the panel of forecasts with Principal Component (PCA) method and then uses the output in a linear regression to obtain the final predictions. 

In this paper,  we describe a novel application of a Factor-QRA method. It extends the PCA forecast averaging approach of \cite{mac:uni:ser:20} to probabilistic forecasting. Moreover it enhances the analysis of \cite{mac:now:wer:16} in three ways. Firstly, we analyze a rich panel of 673 predictions rather than a modest panel of diversified forecasts based on different models. Second, we allow for a dynamic selection of the number of factors with the Bayesian Information Criterion. The results indicate that, unlike in the previous work, the inclusion of a larger number (up to 6) of factors improves the accuracy of the forecast. Finally, the point forecasts that are collected in the panel are subjected to standardization.
In contrast to the predominant use of PCA, the standardization process is implemented across the cross-sectional  rather than the time dimension. It is also applied to the dependent variable, requiring a subsequent transformation of the final outcomes. As a result, it allows to incorporate the impact of point forecast heterogeneity, measured by a forecast variance, on the width of the prediction intervals. Although the outcomes suggest that the standardization has a crucial impact on forecast performance, its significance has not been studied in the literature.

The forecast accuracy of the discussed methods is evaluated with datasets from two European power markets: the German EPEX SPOT and the Polish Power Exchange (TGE). The data for the EPEX day-ahead market covers the tranquil period of 2015-2020, the COVID-19 pandemic and the turbulent period of the Ukrainian war. It encompasses different dynamics of the electricity prices and hence allows for a comprehensive comparison of different forecasting methods. The results show that (i) factor-based averaging schemes provide more accurate probabilistic forecasts than their QRA/QRM counterparts, (ii) the empirical coverage of the new methods is close to the nominal level and passes the \cite{chr:98} test in most cases, (iii) methods using standardized panels are more accurate than their not standardized counterparts.

Finally, to assess the economic value of the proposed forecasting schemes, an experiment is conducted that mimics a real-world trading problem. We consider a decision problem of a medium-sized energy storage unit. The battery operating utility is assumed to offer the electricity on the day-ahead market and therefore it needs to decide about the time of the trade and the price of an offer one day before the delivery. It should be underlined that effective management of batteries becomes an important issue because electricity storage systems are necessary complements of intermittent generation from Renewable Energy Sources (RES).
The results demonstrate that data-driven trading strategies outperform the unlimited-bids benchmark with respect to the average income per 1~MWh. The improvement is associated with a reduction of the volume of trade, which in turn leads to a fall of variable costs and a slowdown of the efficiency degradation process of batteries.


The article is structured as follows. First, we present the data sets, which comprise day-ahead price series and exogenous variables. At the end of Section \ref{sec:data} we specify the data transformation used in the analysis. Next, in Section \ref{sec:point_forecasts} we describe the forecasting model applied to calculate the point predictions. In Section \ref{sec:methodology} we first present the choice of literature benchmarks and then propose the novel approach for the construction of probabilistic predictions. Finally, in Section \ref{sec:res} we present the results of our study and describe the trading strategy used to compare the economic value of the forecasts. The last Section \ref{sec:conclusion} concludes the research.

\section{Data}
\label{sec:data}

\begin{figure*}[b!]
	\centering
	\includegraphics[width = .9\textwidth]{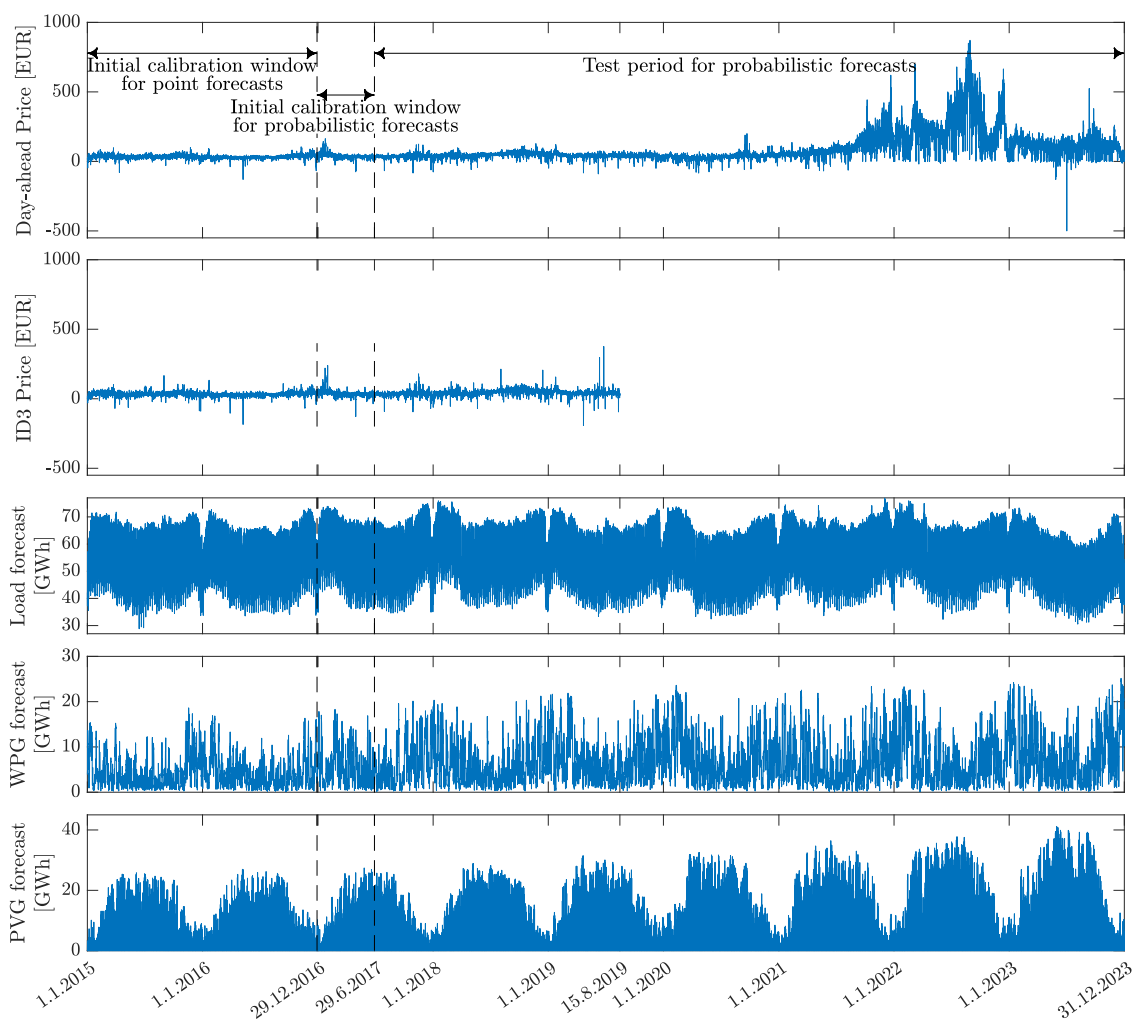}
	\caption{EPEX day-ahead prices (\textit{top}), EPEX ID3 prices (\textit{middle top}), day-ahead load prognosis (\textit{middle}), day-ahead forecasts of wind power generation (\textit{middle bottom}), day-ahead forecasts of Photovoltaic generation (\textit{bottom}) from 1.01.2015 to 31.12.2023 (15.08.2019 for ID3 prices). The first vertical dashed line marks the end of the 728-day calibration period and the second marks the end of the initial 182-day calibration window for probabilistic forecasts. Note that the scales on the y-axes differ.}
	\label{fig:EPEX}
\end{figure*}

\begin{figure*}[tbh]
	\centering
	\includegraphics[width = .9\textwidth]{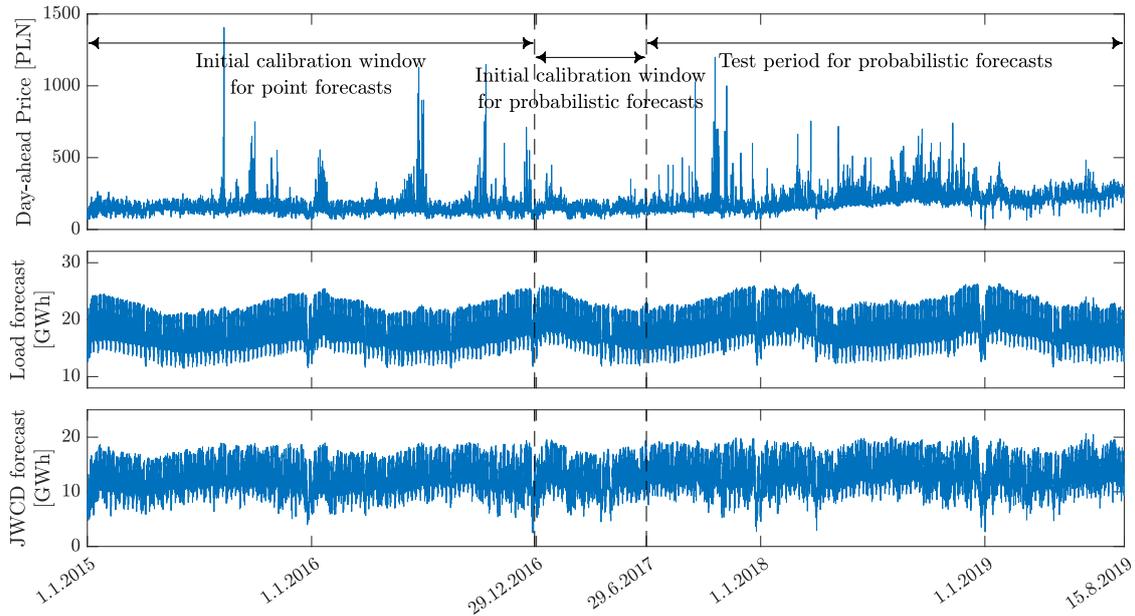}
	\caption{TGE day-ahead prices (\textit{top}), day-ahead load prognosis (\textit{middle}), day-ahead forecasts of Generation of Centrally Dispatched Generating Units (JWCD; \textit{bottom}) from 1.01.2015 to 15.08.2019. The first vertical dashed line marks the end of the 728-day calibration period and the second marks the end of the initial 182-day calibration window for probabilistic forecasts.}
	\label{fig:TGE}
\end{figure*}

To test the methodology proposed in this paper, datasets from two European power markets are considered: the German EPEX and the Polish Power Exchange (TGE). For the German dataset (EPEX), we consider two different price time series: the day-ahead (DA) hourly electricity prices (\emph{top} panel in Figure \ref{fig:EPEX}) and the corresponding time series linked to the intraday market: the hourly prices of the ID3 index (\emph{middle top} panel in Figure \ref{fig:EPEX}). Since electricity on the German intraday market is traded continuously, there is no clear definition of the intraday electricity price. The ID3 index is the most common proxy for the above price and is calculated as the volume-weighted average price of all transactions between 180 and 30 minutes before electricity delivery (see \cite{nar:zie:20JCM}). 
In addition to price series, we consider different types of exogenous variables (see Figure \ref{fig:EPEX}): day-ahead load forecast (\emph{middle} panel) as well as day-ahead prediction of wind (\emph{middle bottom} panel) and solar power generation (\emph{bottom} panel)). Wind generation forecasts consist of the aggregated offshore and onshore generation predictions. All considered series cover the period from 1 January 2015 to 31 December 2023 except for ID3 price which cover period from 1 January 2015 to 15 August 2019.

The second data series comes from the market operated by the Polish Power Exchange (TGE). The dataset, which also covers the period from 1 January 2015 to 15 August 2019, includes three time series with hourly resolution: day ahead prices for fixing no. 1, day-ahead predictions of system load and day-ahead predictions of the so-called Generation of Centrally Dispatched Generating Units (JWCD); see Figure \ref{fig:TGE}. The JWCD term refers to a conventional generating unit connected to the grid subject to central disposition by the transmission system operator.
Note that due to the minimal trading volumes for TGE intraday contracts, contrary to the German data, we do not consider any time series related to the Polish intraday market.

For both markets, we also use two other fundamental indices that influence electricity prices: natural gas spot prices and European Emission Unit Allowance (EUA) spot prices. Unlike electricity and production forecasts, EUAs and natural gas are closing prices and are therefore quoted on a daily (rather than hourly) basis.  In addition, these commodities are traded from Monday to Friday, and therefore the values for weekend days are substituted by the most recent Friday closing price.

All time series were pre-processed to account for changes to/from daylight saving time, as in \cite{wer:06}. Missing values (corresponding to changes in summer time) were replaced by the arithmetic mean of observations from neighbouring hours. The doubled values (corresponding to the changes due to summer time) have been replaced by their arithmetic mean. Due to pronounced spikes and seasonality of electricity prices \citep{jan:tru:wer:wol:13}, we follow \cite{uni:wer:zie:18} and apply the Normal-distribution Probability Integral Transform (N-PIT) to all datasets (to time series of prices and exogenous variables). The transformed variable $Y_{d,h}$ for day $d$ and hour $h$ is given by:
\begin{equation}
\tilde{Y}_{d,h} = N^{-1}\left( \hat{F}_{Y_{d,h}}(Y_{d,h}) \right),
\end{equation}
where $\hat{F}_{Y_{d,h}}(\cdot)$ is the empirical cumulative distribution function of $Y_{d,h}$ in the calibration window, and $N^{-1}$ is an inverse of the normal distribution function.

\section{Point Forecasts}
\label{sec:point_forecasts}

In this study, we consider day-ahead forecasts of electricity prices for both the intraday and day-ahead markets. As they are computed at the same time, they are based on identical information sets and could be used for risk management or trading activities that involve taking positions on both markets (see \cite{mac:nit:wer:21}, \cite{jan:woj:22}).

\subsection{Model for Day-ahead markets}
\label{ssec:DA}

The choice of the benchmark model is motivated by two factors. First, studies on automated variable selection \citep{uni:now:wer:16,wer:zie:18} allow us to optimize the structure of the model and include only the most important predictors. Note that compared to most common structures of the EPF model, we additionally include the price of emission allowances and the price of natural gas, as we believe that they may play an important role in the formation of electricity prices. Second, the computational efficiency required to produce probabilistic forecasts is extremely high. Therefore, in order to minimise the computational effort required to obtain point forecasts, we decide to use a simple autoregressive structure with exogenous variables instead of more complex machine learning techniques.
The model for the transformed day-ahead price on day $d$ and hour $h$ is given by:
\begin{align}
DA_{d,h} = &~\underbrace{ \sum_{i=1}^{7}\beta_{h,11+i} D_{d,i} }_{\scriptsize \mbox{weekday dummies}} + \underbrace{ \beta_{h,1}  DA_{d-1,h} + \beta_{h,2} DA_{d-2,h} + \beta_{h,3} DA_{d-7,h} }_{\scriptsize \mbox{autoregressive effects}} + \underbrace{ \beta_{h,6}  DA_{d-1,24} }_{\scriptsize \mbox{midnight price}} \nonumber \\
& + \underbrace{ \beta_{h,4} DA_{d-1,min} + \beta_{h,5} DA_{d-1,max}}_{\scriptsize \mbox{non-linear effects}}   + \underbrace{ \beta_{h,10} EUA_{d-1} }_{\scriptsize \mbox{emission allowance price}} +  \underbrace{ \beta_{h,11} NG_{d-1} }_{\scriptsize \mbox{natural gas price}} \nonumber \\
&   + \underbrace{ \beta_{h,7} L_{d,h} }_{\scriptsize \mbox{load forecast}} + \underbrace{ \mathds{1}_{8<h<18} \quad \beta_{h,8} S_{d,h} }_{\scriptsize \mbox{solar gen. forecast}} + \underbrace{ \beta_{h,9} W_{d,h} }_{\scriptsize \mbox{wind gen. forecast}}  + \varepsilon_{d,h},
\label{eqn:DA_model}
\end{align}
where $DA_{d-1,h}$, $DA_{d-2,h}$, $DA_{d-7,h}$ include information about autoregressive effects and correspond to prices at the same hour of the previous day, two days earlier, and one week earlier. The choice of autoregressive lags is dictated by seasonalities present in electricity markets \citep{jon:etal:13}. $DA_{d-1,min}$, $DA_{d-1,max}$ and $DA_{d-1,24}$ represent the minimum, maximum, and last known prices of day $d-1$. $L_{d,h}$, $S_{d,h}$ and $W_{d,h}$ refer to the transformed day-ahead load, photovoltaic generation, and wind power generation forecasts for given hour of a day, respectively. Finally, $D_{d,1},..., D_{d,7}$ are weekday dummies and $\varepsilon_{d,h}$ is the noise term. Note that solar generation forecasts, $S_{d,h}$, are included only in the models describing German electricity markets (due to Polish data availability issues). Furthermore, due to lack of generation during night and evening hours, the variable is considered in regressions (\ref{eqn:DA_model})-(\ref{eqn:IDA_model}) only for hours 9-17.

\subsection{Model for EPEX intraday market}
\label{ssec:IDA}

The second model is used to predict the German intraday market price for the next day, computing the forecast on the preceding day $d-1$ (analogously to the DA case). Consequently, intraday and day-ahead prices are forecasted with a very similar methodology: prices for all 24 hours in day $d$ are forecasted simultaneously, using the same pool of information. The model, denoted by $\mathbf{IDA}$ (Intraday Day-Ahead) assumes that the data-generating process of intraday prices could be described by the following equation:

\begin{align}
ID3_{d,h} = & \sum_{i=1}^{7}\beta_{h,12+i} D_{d,i}  + \beta_{h,1}  ID3_{d-1,h}^{*} + \beta_{h,2} ID3_{d-2,h} + \beta_{h,3} ID3_{d-7,h} + \beta_{h,4} DA_{d-1,h}\nonumber\\
&  +  \beta_{h,5}  DA_{d-1,24}+ \beta_{h,6} DA_{d-1,min} + \beta_{h,6} DA_{d-1,max}  +  \beta_{h,11} EUA_{d-1}  \nonumber \\
& +  \beta_{h,12} NG_{d-1} + \beta_{h,7} L_{d,h}  + \mathds{1}_{8<h<18} \quad \beta_{h,9} S_{d,h}  + \beta_{h,10} W_{d,h} + \varepsilon_{d,h},
\label{eqn:IDA_model}
\end{align}


Note that the model (\ref{eqn:IDA_model}) is very similar to the model $\mathbf{DA}$ except that the three autoregressive predictors refer to ID3 prices instead of Day-Ahead and additionally the model includes the Day-Ahead price for the same hour of the previous day. Furthermore, the variable responsible for the price on day $d-1$ is marked with an asterisk, that is $ID3_{d-1,h}^{*}$, because the predictor changes depending on the hour $h$. Due to the fact that forecasting is performed for all hours at the same time, i.e. at 10:00, for later hours the value of the ID3 index for day $d-1$ has not been established yet. Therefore, the value of $ID3_{d-1,h}^{*}$ is the following:

\begin{equation}\label{eqn:partial}
 ID3_{d-1,h}^{*} = 
\begin{cases} ID^{\mathrm{partial}}_{d-1,h} &\mbox{for } h > 10, \\ 
 ID3_{d-1,h} & \mbox{for } h \leq 10,
\end{cases}
\end{equation}
where $ID^{\mathrm{partial}}_{d-1,h}$ is the volume-weighted average price of all transactions for a certain product that have occurred up to the time of forecasting. If there were no transactions, $ID^{\mathrm{partial}}_{d-1,h}$ is replaced by the corresponding day-ahead price.

\subsection{Forecasting framework}
\label{ssec:point}
The model weights in Equations \ref{eqn:DA_model} and \ref{eqn:IDA_model}, $\boldsymbol{\beta_{h}}$, are estimated by minimising the residual sum of squares (RSS), independently for each hour in the out-of-sample period (see Section \ref{sec:data}). We consider a rolling calibration window scheme, but rather than arbitrarily choosing a fixed calibration window length, we consider data samples ranging from 56 to 728 days. By $\hat{P}_{d,h}(\tau)$ we denote the forecast for day $d$ and hour $h$ estimated using the $\tau$ day calibration window. For each day and hour, we obtain 673 different forecasts and, as shown in recent studies \citep{ser:uni:wer:19}, the information contained in such a rich panel of forecasts can lead to predictions that are statistically superior to any single-window-based forecast. Furthermore, the performance of these forecasts tends to be more consistent across different data sets \citep{mac:uni:ser:20}. The point forecasts obtained from the autoregressive models form the basis of all the probabilistic models considered in this paper.

\section{Probabilistic forecasts}  
\label{sec:methodology}

In recent years, probabilistic forecasting has gained attention in the EPF literature, as it complements point predictions with additional information about potential trade risk. In the following sections, we describe different approaches for obtaining probabilistic forecasts. They are all calculated using data from the second calibration window (see Figure \ref{fig:EPEX}), which includes 182 days of observations (half a year). The choice of the length of this sample is inspired by the results of \cite{ser:uni:wer:19} and should provide enough data points for stable estimation of the model parameters. In methods based on Quantile Regression Averaging, we use as input point predictions, $\hat{P}_{d,h}(\tau)$, derived from regressions (\ref{eqn:DA_model}) - (\ref{eqn:IDA_model}). When historical simulations and conformal predictions are considered, we explore the out-of-sample point forecast errors, $\hat{\varepsilon}_{d,h}$, calculated as
\begin{equation*}
    \hat{\varepsilon}_{d,h} = P_{d,h} - \hat{P}_{d,h}.
\end{equation*}
Point forecast $\hat{P}_{d,h}$ is computed as an average of predictions obtained with six different calibration windows $\tau_i\in\{ 56, 84, 112, 714, 721, 728\}$:
\begin{equation*}
    \hat{P}_{d,h} = \sum_{i=1}^6\hat{P}_{d,h}(\tau_i).
\end{equation*}
 The selection of window sizes, $\tau_i$ is based on the results of \cite{hub:mar:wer:19} and \cite{ser:uni:wer:19}.

In this paper, we assume that the distribution of future electricity prices can be approximated by a range of prediction intervals ($PI$s). The intervals are characterized by a nominal coverage $1-\alpha$. It is assumed that for a given level of coverage, the $PI_{1-\alpha}$ contains the future value of the dependent variable with a probability $1-\alpha$. At the same time, the observation falls outside of the $PI_{1-\alpha}$ with probability $\alpha$. Hence, $\alpha$ can be interpreted as an uncertainty level and impacts the range and the width of $PI$s.

\subsection{Historical simulations}
\label{ssec:his}
Historical simulation is a direct method of constructing probabilistic forecasts. It takes different names in the literature and sometimes is called an empirical error distribution approach \citep{kat:zie:18}. It explores both, the point forecasts and forecast errors. It is constructed as follows:
\begin{equation}
        PI_{1-\alpha}^{hist} = [\hat{P}_{d+1,h}+\gamma_{\alpha/2}, \hat{P}_{d+1,h}+\gamma_{1-\alpha/2}],
\end{equation}
where $1-\alpha$ is the $PI$'s nominal coverage level. Variables $\gamma_{\alpha/2}$ and $\gamma_{1-\alpha/2}$ describe the $(\alpha/2)$ and the $(1-\alpha/2)$ quantiles of $\hat{\varepsilon}_{d,h}$, respectively.
\subsection{Conformal predictions}
\label{ssec:cp}

The concept of  conformal predictions was originally introduced by \cite{vovk:gam:vap:98} and \cite{vovk:etal:09} and later developed by \citep{ vovk:etal:18, vovk:etal:19,  wan:etal:23}. The basic algorithm uses a conformity score to divide a set of future values into those inside and outside of a prediction interval. Unfortunately, it requires re-estimation of model parameters for each potential value of the dependent variable and hence is computationally burdensome. \cite{lei:etal:18} and \cite{kat:zie:21} proposed an alternative way of constructing prediction intervals, which requires fewer computations and, hence, can be used in complex modeling frameworks. First, the out-of-sample point-prediction errors are computed by, for example, a sample splitting. Next, the non-conformity score, here the absolute value of the error, is calculated. Finally, $PI_{1-\tau}^{conf}$ is constructed as
\begin{equation}
    PI_{1-\alpha}^{conf} = [\hat{P}_{d+1,h}-\lambda_{\alpha}, \hat{P}_{d+1,h}+\lambda_{\alpha}],
\end{equation}
where the parameter $\lambda_{\alpha}$ is calculated as the $\alpha$-th quantile of $|\hat{\varepsilon}_{d,h}|$.

\subsection{Quantile Regression Averaging}
\label{ssec:qr}

Quantile regression (QR) is a general approach, which allows to represent a quantile of order $q$ of a dependent variable, here $\hat{P}^{q}_{d,h}$, as a linear function of a set of inputs $X_{i,d,h}$:
\begin{equation}
	\hat{P}^{q}_{d,h} = w_{0,q} + \sum_{i=1}^K w_{i,q}X_{i,d,h}.
	\label{eq:QR}
\end{equation}
The parameters $w_{0,q},..., w_{K,q}$, are estimated by minimizing the so-called \emph{pinball score}, given by:
  \begin{equation*}
    PS_{q}= \begin{cases}
 (1-q)(\hat{P}^{q}_{d,h} - P_{d,h})  & \text{ for } P_{d,h} < \hat{P}^{q}_{d,h}, \\
 q(P_{d,h} - \hat{P}^{q}_{d,h}) & \text{ for } P_{d,h} \geq \hat{P}^{q}_{d,h}.
 \end{cases}
 \end{equation*}
The estimation process can be carried out separately for 99 percentiles: $q=0.01,...,0.99$ and hence allows us to approximate the whole distribution of $P_{d,h}$.

In recent years, QR has been used successfully as a method of averaging forecasts. \cite{now:wer:15} proposed using multiple point forecasts as input in (\ref{eq:QR}) and showed that such a Quantile Regression Averaging (QRA) approach resulted in more accurate interval forecasts compared to individual models. The idea was subsequently explored and developed by several authors. \cite{mar:uni:wer:19:narx} proposed a modification called the Quantile Regression Machine (QRM), which first applies a simple mean to the average of the point forecasts and then uses the improved forecast as a single input to QR. The above two forecast averaging schemes have been compared by \cite{ser:uni:wer:19}, who indicate that QRM outperforms QRA when averaging is applied to predictions derived from a single model calibrated to windows of different lengths.

In this research, two QRA benchmarks are considered. First, the selected point predictions obtained with six different calibration windows, $\tau = 56, 84, 112, 714, 721, 728$, are used directly as explanatory variables in the quantile regression.
\begin{equation}
	\hat{P}^{q}_{d,h} = w_{0,q} + \sum_{i=1}^6 w_{i,q}\hat{P}_{d,h}(\tau_i).
\end{equation}

Secondly, the same set of point forecasts are first combined to obtain a averaged prediction, a then it is used as the only input in the quantile regression:
\begin{equation}
	\hat{P}^{q}_{d,h} = w_{0,q}+w_{1,q}\hat{P}_{d,h}.
\end{equation}
The two benchmarks are denoted QRA and QRM, respectively. The visualizations of those two benchmarks, denoted QRA and QRM, respectively, are depicted in the first two rows of Figure \ref{fig:flowchart}. QRA and QRM methods can be directly used for constructing prediction intervals. In order to create a $PI$ of the nominal coverage $1-\alpha$, forecasts of two quantiles $\alpha/2$ and $1-\alpha/2$ are used:
\begin{equation}
\label{eq:PI}
        PI_{1-\alpha} = [\hat{P}^{\alpha/2}_{d+1,h}, \hat{P}^{1-\alpha/2}_{d+1,h}].
\end{equation}

Note that both methods use only six preselected point forecasts. This model specification is inspired by \cite{hub:mar:wer:19}, who showed that a combination of three short and three long windows yields the largest gains in forecast accuracy. Furthermore, \cite{ser:uni:wer:19} showed that including a larger set of point predictions in the QR can lead to a decrease in prediction accuracy. There are two main reasons for such a decrease: (i) as the number of inputs increases, parameters are estimated with a larger error, (ii) individual point forecasts are highly correlated, so new inputs add little or no additional information to the model.

\subsection{Factor Quantile Regression Averaging}
\label{ssec:fqr}

Although only six point forecasts are used in the QRA and QRM methods, all forecasts from calibration windows ranging from 56 to 728 days form the basis of the next group of probabilistic models. To explore the full panel of 673 point forecasts, which are highly correlated and therefore should not be used together in a regression model, similar to \cite{uni:mac:22}, we propose to use Principal Component Analysis (PCA). PCA allows a large number of variables to be represented by a few principal components, called factors, which are assumed to be orthogonal and therefore solve the colinearity problem mentioned above.

Being one of the most popular methods of dimension reduction technique \citep{vel:19,he:21,guo:22}, PCA has been shown to be useful in the construction of probabilistic forecasts. \cite{mac:now:wer:16} proposed to apply quantile regression to the first, most prominent factor and to call it Factor Quantile Regression Averaging (FQRA). Based on empirical studies, the authors conclude that FQRA outperforms standard QRA. In this paper, PCA is applied to a relatively small panel of predictions from 32 different models/model specifications. 

In this paper, we explore a slightly modified approach. First, probabilistic forecasts are computed jointly for all hours of a day with a single model, rather than with a separate model for each hour. Thus, electricity prices and their forecasts are treated as time series with a time index $t=24(d-1)+h$ This setup allows the inclusion of information about the underlying correlation between prices in neighboring hours. Since we extract factors for both past and current data, the analyzed window is extended by additional $24$ forecasts of hourly prices from the forecast day, $d_f$. Let $\hat{P}_{t}(\tau)$ denote the forecast of the variable $P_t$ based on a calibration window of length $\tau$. The data set
$\hat{P} =\{\hat{P}_{t}(\tau)\}$ constitutes a $(T\times N)$ panel, {with $T=24(182+1)$ and $N=673$. The first dimension of the panel represents the time, and the second dimension describes the size of a calibration window.

\begin{figure*}[t]
	\centering                   
\includegraphics[width = .85\textwidth]
{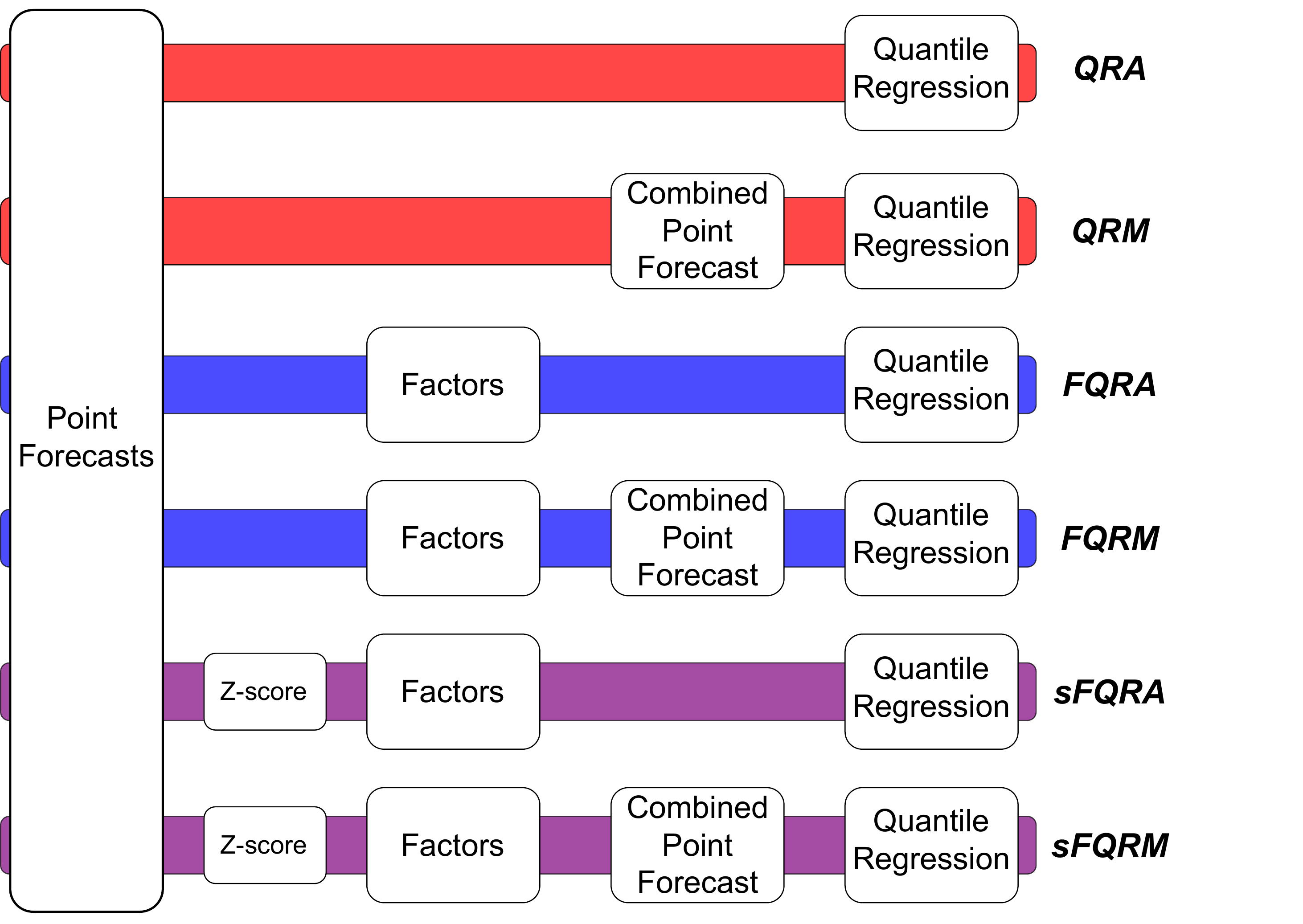}
	\caption{The flowchart illustrating the subsequent steps of forecasting models.}
	\label{fig:flowchart}
\end{figure*}

All PCA-based models are based on common factors extracted from the panel of point forecasts $\hat{P}$. The factors, $F_{t,k}$ for $k=1,...,K$, are calibrated with the Least Squared method described by \citep{sto:wat:02a, bai:ng:02}. They are estimated as $\sqrt{T}$ times the eigenvectors corresponding to the $K$ largest eigenvalues of the $(T\times T)$ matrix $\hat{P}\hat{P}'$. Alternatively, they can be obtained as $\sqrt{N}$ times the eigenvectors of the $(N\times N)$ matrix $\hat{P}'\hat{P}$.

Factors are used in two ways in the process of obtaining price percentile forecasts. They are either used first to obtain the point forecast (as in \cite{mac:uni:ser:20}), which is then utilized as input for the quantile regression, or they are used directly as regressors for the quantile forecast estimation. We denote the first approach as FQRM/sFQRM and the second as FQRA / sFQRA, corresponding to the QRM and QRA methods discussed in Section \ref{ssec:qr} (see Figure \ref{fig:flowchart}). The detailed regression equations for these methods are given below. In the FQRA/sFQRA models, quantile regression is used as the link between the factors summarizing the point forecasts and the quantile of the electricity price:
\begin{equation}
P^{q}_{t} = w_{0,q} + \sum_{k=1}^K w_{k,q}\hat{F}_{t,k}
\label{eq:FQRA}
\end{equation}
where $\hat{F}_{t,k}$ is the estimator of a $k$-th factor at the time $t$, with $k=1,...,K$. Similarly to the QRM approach, in the FQRM/sFQRM methods, the point forecast is first calculated using a linear regression.
\begin{equation}
	P_{t} = \beta_0 + \sum_{k=1}^K \beta_k\hat{F}_{t,k} + e_t.
	\label{eq:PCA}
\end{equation}
Next, the fitted values $\hat{P}_t$ obtained with the above model are used in a quantile regression
\begin{equation}
	\hat{P}^{q}_{t} = w_{0,q} +  w_{1,q}\hat{P}_t,
    \label{eq:FQRM}
\end{equation}
In both cases, for regressions (\ref{eq:FQRA}) and (\ref{eq:PCA}), Bayesian Information Criterion (BIC) is used to select the optimal number of factors, $K$. 

Finally, we consider two specifications of the FQRA and FQRM approaches that depend on a standardisation of the input data. In the FQRA/FQRM methods, the factors are estimated directly from the panel $\hat{P}$. In sFQRA and sFQRM approaches, on the other hand, the data is first standardised according to the following formula:

\begin{equation}
     \hat{p}_{t}(\tau)=\frac{\hat{P}_{t}(\tau)-\hat{\mu}_{t}}{\hat{\sigma}_{t}}, \nonumber
\end{equation}
where $\hat{\mu}_{t}$ is the mean and $\hat{\sigma}_{t}$ is the standard deviation of forecasts $\hat{P}_{t}(\tau)$ across different window sizes, $\tau$. The standardization step is denoted as \textit{Z-Score} in Figure \ref{fig:flowchart}. Next, the panel $\hat{p} = \{\hat{p}_{t}(\tau)\}$ is used to estimate factors, $F_t$. Whenever the standardization is used, it is also applied to the predicted variable
     \begin{equation}
         p_{t}=\frac{P_{t}-\hat{\mu}_{t}}{\hat{\sigma}_{t}}. 
     \end{equation} 
It should be noted that $\hat{p}$ and $p$ have different units than $\hat{P}$ and $P$. Therefore, once the quantile predictions $\hat{p}^{\alpha}_{t}$ are calculated with Equations (\ref{eq:FQRA}) or (\ref{eq:FQRM}), they are transformed back into the original units
    \begin{equation}
    \label{eq:transf_back}
        \hat{P}^{q}_{t}=\hat{p}^{q}_{t}*\hat{\sigma}_{t}+\hat{\mu}_{t}.
     \end{equation}
Empirical analysis shows that standardization has a significant impact on the performance of the proposed averaging schemes.

Tu sum up, Factor Quantile Regression averaging method consists of the following steps (depicted on Figure \ref{fig:flowchart}):
\begin{enumerate}
    \item The panel $\hat{P}=\{\hat{P}_{t}(\tau)\}$ is constructed from a set of point forecasts $\hat{P}_{d,h}(\tau)$. In the sFQRA and sFQRM approaches, the data is further standardized and denoted by $\hat{p}$.
    \item $K$ factors summarizing the information in the panel $\hat{P}$ or $\hat{p}$ are estimated using the PCA method.
    \item The estimated factors, $\hat{F}_{t,k}$, are used to compute probabilistic forecasts with quantile regression using equations (\ref{eq:FQRA}) or (\ref{eq:PCA})-(\ref{eq:FQRM}). The weights in the equations are estimated using the last 182 days preceding the forecast day.
    \item In the sFQRA and sFQRM methods, the predictions are transformed into the original units with (\ref{eq:transf_back}).
    \item The prediction intervals, $PI_{1-\alpha}$, are constructed according to (\ref{eq:PI}).
\end{enumerate}

\section{Results}
\label{sec:res}

\subsection{Statistical measures}


In this research, the distribution of future values of electricity prices is approximated by prediction intervals.
The competing forecasting schemes are first compared on the basis of their empirical coverage \citep{cha:93}. For each day and hour of the testing period, we calculate the 'hits' of the prediction intervals:
\begin{equation}
I^{1-\alpha}_{d,h} = 
    \begin{cases}
        1 & \text{if } P_{d,h} \in PI_{1-\alpha}\\
        0 & \text{if } P_{d,h} \not\in PI_{1-\alpha}\\
    \end{cases}
\end{equation}
Next, the average number of 'hits' is calculated, which represents an empirical coverage level. For a given hour $h$, it is computed as
\begin{equation}
Cov^{1-\alpha}_h= \frac{1}{N}\sum_{d} I^{1-\alpha}_{d,h},
\end{equation}
where $N = 778$ (for German ID3 and Polish day-ahead) or $N=2377$ (for German DA) is number of days in out-of-sample period. The average coverage over all hours is defined as:
\begin{equation}
Cov_{1-\alpha}= \frac{1}{24}\sum_{h} Cov^{1-\alpha}_h.
\end{equation}

Next, the hypothesis $H_0: Cov^{1-\alpha}_h=1-\alpha$ is tested with the Christoffersen's test \citep{chr:98} for each hour separately. The test takes into account not only the unconditional coverage of the prediction intervals ($Cov^{1-\alpha}_h$), but also the independence of the exceedances of the quantile levels in successive time periods. This particular property of probabilistic forecasts is often overlooked in the literature, as most studies focus only on the unconditional coverage of prediction intervals. Christoffersen's test can be seen as an extension of a popular Kupiec test \citep{kup:95}.

\subsection{Results}


We consider 25 levels of coverage ranging from $1-\alpha = 50\%$ to $98\%$ to compare the predictive accuracy of the proposed methods. 
In Figure \ref{fig:coverages} we report the average coverage error \citep[ACE;][]{now:wer:18}, $Cov_{1-\alpha}-(1-\alpha)$, for three markets and eight models analyzed in this study. Note that for accurate $PI$ predictions, the ACE will be close to zero. The results lead to the following conclusions:

 \begin{figure}[p]
     \centering
      \includegraphics[width= .9\textwidth]{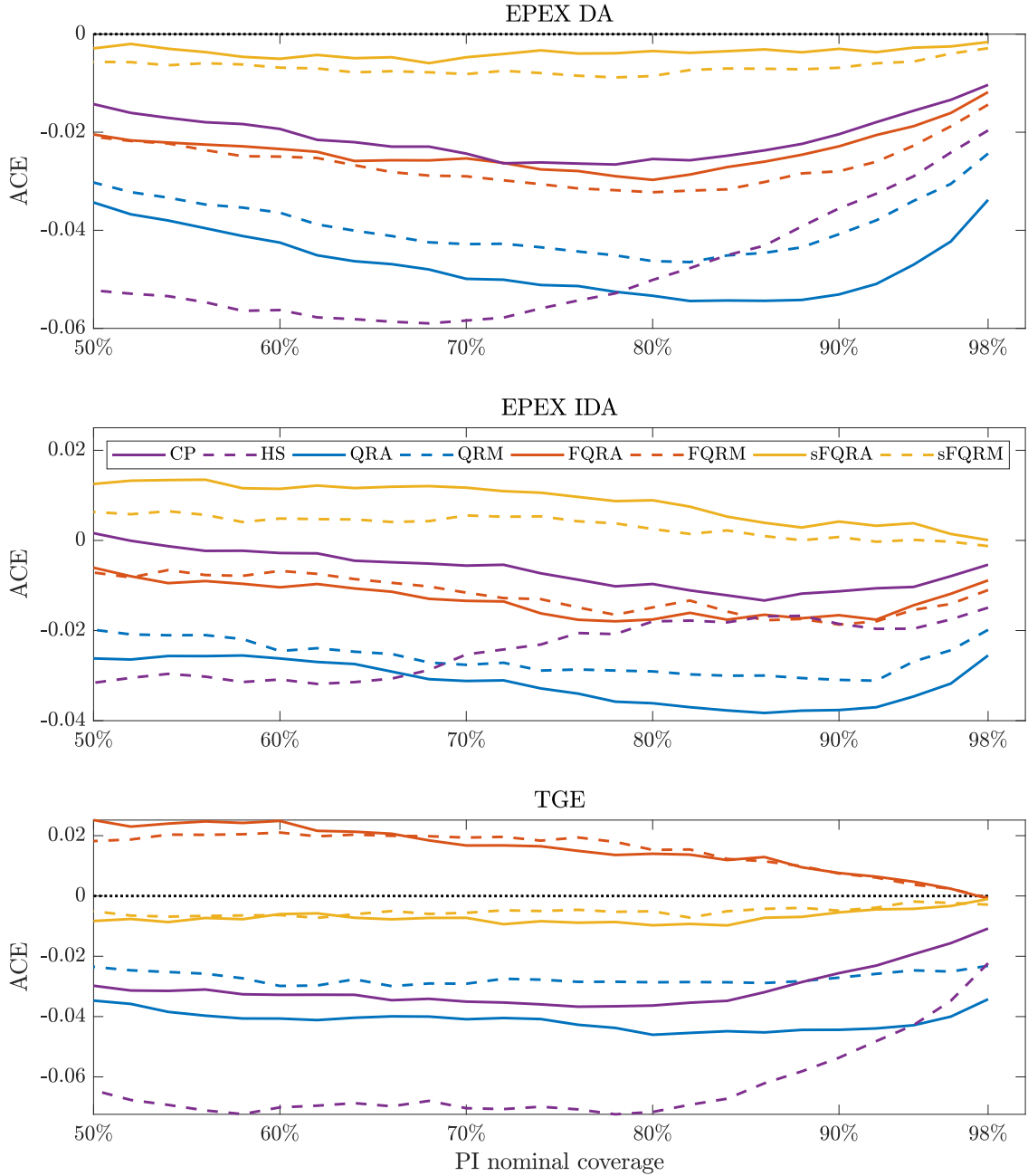}
    \caption{Results of the average coverage error for prediction interval levels ranging from $1-\alpha = 50\%$ to $98\%$ obtained with all considered probabilistic forecasting methods. The results are divided into three panels for the German Day-ahead, German Intraday and Polish Day-ahead markets (consecutively from top to bottom).}
	\label{fig:coverages}
\end{figure}

\begin{itemize}
    \item When empirical coverage is considered, the sFQRA and sFQRM methods clearly outperform their competitors. For both German and Polish day-ahead markets they provide predictions with the empirical coverage closest to the nominal level for nearly all considered prediction intervals.

      \item Forecasts from HS, CP and the QR-based (QRA and QRM) models provide PIs that are on average too narrow and show consistently lower coverage for all datasets and nearly all values of nominal coverage. On the other hand, the hybrid PCA-QR approaches produce wider PI for German Intraday (FQRA/FQRM) and Polish day-ahead markets (sFQRA/sFQRM).
    
    

\end{itemize}

Table \ref{tab:coverage_test} shows the number of hours (out of 24), for which the null of the Christoffersen test was not rejected for three selected coverage levels $1-\alpha = 50\%, 80\%$ and $98\%$. This indicates that the empirical coverage is close enough to the nominal level and that the quantile exceedances are independent. Based on the results, we can conclude that:
\begin{itemize}
    \item The forecasts from the sFQRA and sFQRM models pass the Christoffersen test for the largest number of hours for each nominal coverage value and all datasets. 
    \item None of the methods pass the test for 50\% and 80\% prediction interval for the German day-ahead. It seems to be related to the huge increase in volatility of the prices in years 2021-2023, which resulted in poor accuracy of the underlying point forecasts.
    \item The performance of the FQRA/FQRM and QRA/QRM models differs across the nominal coverage levels. The latter models pass the Christoffersen test for a slightly higher number of hours for the 50\% and 80\% levels. However, for 98\% nominal coverage, the forecasts from the FQRA and FQRM models outperform both QR-based methods by a large margin.
    \item The CP benchmark, unlike historical simulation, provides forecasts that pass the Christoffersen test for a relatively high number of hours.
\end{itemize}

In summary, statistical measures show that sFQRA and sFQRM forecast averaging methods outperform their FQRA/FQRM and QRA/QRM counterparts as well as the considered benchmarks. They help to construct PIs for which the empirical coverage deviates the least from its nominal level. Note that at the same time, FQRA/FQRM methods produce results that, despite their increased computational complexity, are not much better than those of QRA/QRM approaches. This underlines the importance of standardization in the estimation of averaging weights.

\begin{table*}[tbp]
	\centering
 \scalebox{0.85}{
	\begin{tabular}{| c c|c|c|c|c|c|c|c|c|c|}
		\hline
\multicolumn{2}{|c|}{Coverage} & \multicolumn{3}{|c|}{ 50\%} &\multicolumn{3}{|c|}{ 80\%} &\multicolumn{3}{c|}{ 98\%} \\
\cline{1-11}
\multicolumn{2}{|c|}{Market} & \multicolumn{2}{|c|}{EPEX} & TGE & \multicolumn{2}{|c|}{EPEX} & TGE & \multicolumn{2}{|c|}{EPEX} & TGE \\[2pt]
\cline{1-11}
\multicolumn{2}{|c|}{Price} & DA	& IDA & DA & DA	& IDA & DA & DA	& IDA &  DA \\[2pt]
\hline
\multicolumn{2}{|r|}{CP} &0 &19 &1 & 0 & 3 & 1 &  0 & 20 & 4\\
\multicolumn{2}{|r|}{HS} &0 &9 &0 & 0 & 1 & 0 &  0 & 7 & 0\\
	 \multicolumn{2}{|r|}{QRA} &0 &14 &6 &  0 & 2 & 3 &  0 & 0 & 1\\
      \multicolumn{2}{|r|}{QRM} &0 &17 &6 &  0 & 2 & \bfseries{4}&  0 & 1 & 4\\
      \multicolumn{2}{|r|}{FQRA}&0 &3 &4 &  0 & 0 & 0  & 1 & 11 & 4\\
      \multicolumn{2}{|r|}{FQRM}&0 &2 &3&  0 & 0 & 0  & 0 & 9 & 8\\
      \multicolumn{2}{|r|}{sFQRA}&0 &\bfseries{20} &\bfseries{9} & 0 & \bfseries{17} & \bfseries{4} & \bfseries{19} & \bfseries{23} &	\bfseries{19}\\
	 \multicolumn{2}{|r|}{sFQRM}&0 &19 &6 & 0 & \bfseries{17} & 3 & \bfseries{19} & 22 & \bfseries{19}\\[3pt]
      \hline
	\end{tabular}
 }
 \caption{Results of the conditional Christoffersen test for selected PI level The table shows the number of hours of the day (out of 24) for which the null hypothesis of the Christoffersen test is not rejected at the 5\% significance levels. The results are divided into three sub-tables for the prediction intervals of $1-\alpha = 50\%, 80\%,$ and $ 98\%$ (consecutively from left to right).}
 \label{tab:coverage_test}
\end{table*}    

\section{Economic evaluation}
The vast majority of articles focus on the evaluation of forecasts and the comparison of predictive models based on statistical measures. Although popular, this approach possesses a number of drawbacks, one of the most important being the choice of the evaluation measure. As \cite{kol:20} argues, the term 'best forecast' strongly depends on the choice of the evaluation measure. Therefore, in order to properly assess the forecasts in the objective manner, a universal measure should be introduced. Evaluating forecasts using an economic measure, such as profit, not only provides a universal approach applicable to any forecast (generated by minimizing any given loss function), but also provides valuable information about the real-world utility of the generated forecasts.

Although only a few articles in the EPF literature consider economic measures for the evaluation of forecasts and the classification of predictive models, the topic has recently started gaining the attention of researchers in the field of EPF. The basis for the economic evaluation of forecasts is the trading strategy that mimics the actual behavior of market participants such as speculators, energy producers or energy consumers. Among the strategies considered in the literature, many involve trading electricity in the day-ahead market or in continuous intraday markets \citep{mac:nit:wer:21,uni:wer:21,ser:mar:wer:22}. Several authors consider a one-sided approach, taking the perspective of the supplier or consumer \citep{zar:can:bha:10,doo:amj:zar:17,jan:woj:22}, while others propose a trading strategy that includes an electricity storage system \citep{kat:zie:18,uni:wer:21,uni:22}.

\subsection{Quantile-based trading strategies}
\label{sssec:Quantile-based:strategies}

Accurate forecasting proves its worth when it translates into financial gains. In this section, we assess the economic significance of probabilistic forecasting through an evaluation of the trading strategy described in \cite{uni:22}. The author presents a trading strategy designed for market participants with access to storage capacity. The study considers a practical scenario in which a company owns a 2.5 MW battery with an efficiency of 90\% for each charge and discharge cycle. Technical constraints prevent the battery from discharging below 20\% of its nominal capacity (i.e., 0.5 MW). The strategy is to buy energy and charge the battery during low price periods, typically in the early morning hours, and then discharge and sell energy during high price periods, typically in the afternoon.

To identify the optimal buying ($h1$) and selling ($h2$) hours and prices, we use probabilistic forecasts derived from the considered methods. First, the optimizer selects the hours with the lowest and highest prices of the day based on the point forecast $\hat P_{d,h}$:

\begin{equation}
	\max_{h1, h2} \left( 0.9~ \hat P_{d,h2} - \frac{1}{0.9}\hat P_{d,h1} \right).
\end{equation}

Subsequently, as shown in Figure \ref{fig:strategy} the company places a bid to buy $\frac{1}{0.9}$ MW at the high (e.g. 95th; $\hat P^{95\%}_{d,h1}$) price percentile forecast for hour $h1$ and simultaneously offers to sell 0.9 MW at the low (e.g. 5th; $\hat P^{5\%}_{d,h2}$) percentile forecast for hour $h2$. In cases where both offers are accepted on the day-ahead market the daily profit equals $0.9P_{d,h2} - \frac{1}{0.9} P_{d,h1}$. If one of the offers is rejected an additional price-taker transaction is executed at the beginning of the following day. Finally, if neither offer is accepted the trader takes no action. For a more comprehensive description of the trading strategy, please refer to \cite{uni:22}.

In this paper we follow a price-taker trading model, i.e. we neglect the market impact of orders placed in the market. Since the strategy involves buying or selling only 1MW we see it as a viable assumption. However, when trading larger volumes, the market impact should be considered (see \cite{nar:ze:22}).

\begin{figure}[tb]
	\centering 
	\includegraphics[width = .65\textwidth]{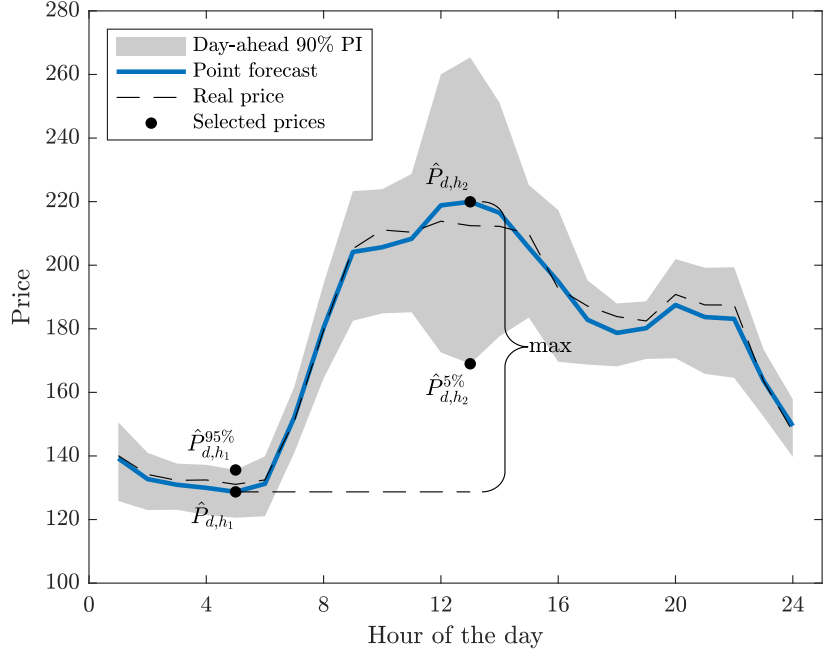}
	\caption{An example of the quantile-based trading strategy of \cite{uni:22} for Polish market on 06.07.2017. Red line indicates the point forecast ($\hat P_{d,h}$) and the dashed black line represents the real price. The grey space depicts the day-ahead forecast of the 90\% prediction interval. Trading hours are selected based on the maximum difference in median forecasts; dots at the PI limits represent the offer $\hat P^{5\%}_{d,h_2}$ and bid $\hat P^{95\%}_{d,h_1}$ prices. }
	\label{fig:strategy}
\end{figure}

\subsection{Unlimited-bids benchmark}
\label{sssec:benchmark:strategies}
The profits made with strategies using probabilistic forecasts are compared with a simple unlimited-bid strategy. Here, the trader selects the moment of a transaction by examining the point forecasts and choosing the hours corresponding to the lowest ($h_1$) and highest ($h_2$) predicted prices. The price-taker offer is then submitted to sell at hour $h_2$ and the price-taker bid is submitted to buy at hour $h_1$. The daily profit that takes into account the efficiency of the battery is calculated as $0.9P_{d,h_2} - \frac{1}{0.9}P_{d,h_1}$.

\subsection{Results}

In this paper, similar to \cite{uni:22}, we compare quantile-based trading strategies for different prediction interval levels ranging from 50\% to 98\% and different forecasting models. Since trading volume may differ conditional on the choice of the model and the prediction interval level, we report profits, defined as the average profit per 1 MWh of trade. Since market participants face different costs associated with trading activities (operational and maintenance costs of batteries, replacement costs, etc.) this measure allows one to easily determine the profitability threshold and avoid trade when the variable costs exceed the income level. Additionally, we report the average traded volume, which shows how often trades are executed for each of the discussed strategies.

The considered testing window, which spans from 29.06.2017 to 31.12.2023, contains  periods of different dynamics of the electricity prices. 
The most significant shift in price behavior was observed in the second half of 2021, when volatility rapidly increased due to rising fuel prices as a result of the Russian invasion on Ukraine. Another significant change in price dynamics occurred approximately at the beginning of 2023, when prices started to stabilize again. Because the profits obtained with the proposed quantile-based trading strategies depend strongly on the level and volatility of prices, the results for the German EPEX market are reported separately for three regimes:
\begin{itemize}
    \item low-level and low-volatility period: from 29.6.2017 to 31.12.2020,
    \item high-level and high-volatility period: from 1.1.2021 to 31.12.2022,
    \item midium-level and midium-volatility period: from 1.1.2023 and 31.12.2023.
\end{itemize}

 \begin{figure}[p]
     \centering
      \includegraphics[width= .9\textwidth]{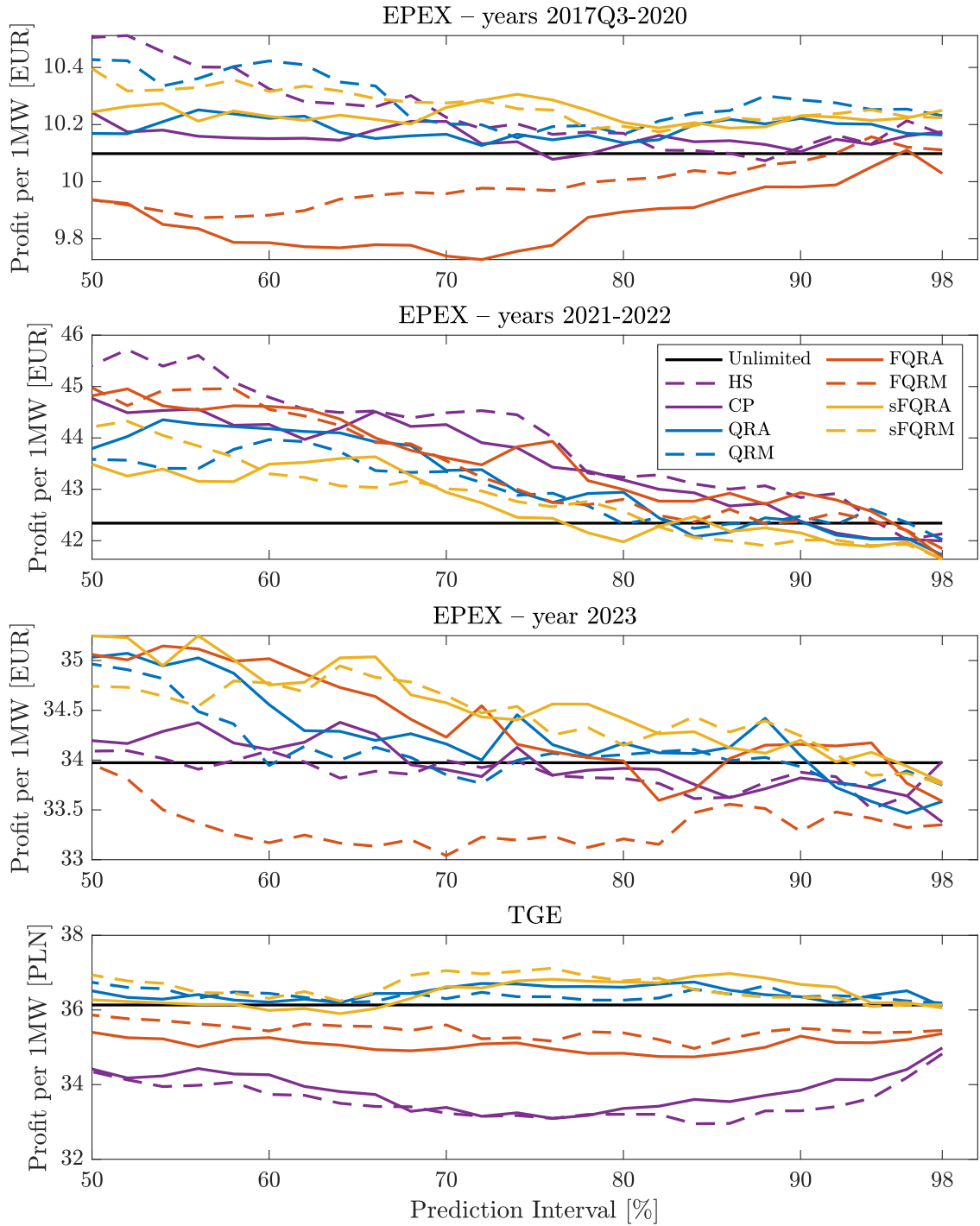}
    \caption{Profits per 1 MWh traded for EPEX SPOT (in Euro; top three panels) and TGE (in PLN; bottom panel). 
    The purple lines refer to the profits obtained with benchmarks (HS -- dashed line, CP -- solid line), blue lines refer QRA (dashed line) or QRM (solid line), the blue red represents FQRA (dashed line) and FQRM (solid line) and the yellow lines represent sFQRA (dashed line) and sFQRM (solid line). The results of the unlimited bidding strategy are plotted with the black solid line.}
	\label{fig:strategy_profits}
\end{figure}

Figure \ref{fig:strategy_profits} shows the average profit per each 1 MWh of trade for EPEX and TGE markets. Calculations do not take into account any financial costs (such as transaction costs or capital costs) other than the loss due to limited charge and discharge efficiency. The results lead to the following conclusions:

\begin{itemize}
    \item Overall it is not a trivial task to indicate the best-performing model as their performance vary across different price volatility regimes and markets.
    \item In low- and mid-volatility price regimes (2017Q3-2020 and 2023) the highest profits from the trading strategies are achieved using the probabilistic forecast from the sFQRA/sFQRM and QRA/QRM models. This is true both for EPEX and TGE marktes. 
    \item For the EPEX market in 2017Q3-2020 and 2021-2022 periods, the HS benchmark is an excellent performer. However, for year 2023 as well as for the TGE market, it is consistently outperformed by the competitors.
    \item The FQRA/FQRM models exhibit the best performance in the high-volatility price regime (2021-2022), whereas they underperform all other approaches in periods with more stable price behavior.
    \item The value of probabilistic forecasts is clearly leveraged for the EPEX market. However, for the TGE market, only probabilistic forecasts from sFQRA/sFQRM and QRA/QRM models allow achieving higher profits than the point forecast-based approach, i.e. the unlimited-bids benchmark.
    \item Interestingly, the model performance in terms of the coverage accuracy (see Figure \ref{fig:coverages}) does not seem to have a clear relation to the economic value of produced forecasts, i.e. the best forecasts in terms of the statistical measures do not have to be the best in terms of the economic evaluation.
\end{itemize}

In Table \ref{tab:stategy:EPEX}, we show traded volumes for out-of-sample periods that correspond to the outcomes presented in Figure \ref{fig:strategy_profits}. The values are expressed relative to the results of the unlimited-bids strategy. It should be recalled here that this benchmark strategy ensures 2~MWh of trade on every day. At the same time, data-driven strategies restrict the trade, when the purchase price exceeds a given threshold or when the sell price falls short of the selected quantile. Therefore, the average volume relative to the benchmark never exceeds 100\%.
 For the sake of readability, we present the results for three selected PI levels (50\%, 80\%, and 98\%). Due to the monotonicity of quantiles, the traded volumes increase with the PI level. Hence, the results for intermediate PIs stay between the reported values.  
 The results lead to the following conclusions:

\begin{itemize}
    
    \item The unlimited-bids strategy trades 2 MW of energy every day, so all other strategies cannot have a volume larger than the benchmark. However, due to the additional price-taker transaction, the total volume traded for quantile-based strategies is not less than 90\%, even for 50\% PI.

    \item QRA based forecasting methods permit more trade than the historical simulation and CP approaches. At the same time, except for the most turbulent times in years 2021-2022, they provide higher income per trade. Therefore, the methods are more effective in detecting low-income periods.

\end{itemize}

To conclude, the results indicate that the data-driven strategies allow restricting the trade in the least profitable periods. As a result, a reduction in trade volume of up to 10\%  leads to a growth of the profit per 1~MWh  of trade. It can be noticed that a lower volume of trade is associated with smaller operational costs and a slower reduction of battery efficiency.
Hence, the data-driven approaches, when properly used, bring higher income at lower variable costs and, in this context, dominate the benchmark unlimited bids approach.

\begin{table}[tbp]
\centering
\caption{Traded volume relative to unlimited-bids benchmark (in percentage) for EPEX SPOT and Polish TGE obtained using a range of quantile-based strategies and all considered probabilistic forecasting models.}
\label{tab:stategy:EPEX}
\scalebox{0.85}{
\begin{tabular}{|r|c|c|c|c|c|c|c|c|c|c|c|c|}
\hline
Market& \multicolumn{9}{c|}{EPEX} & \multicolumn{3}{c|}{TGE} \\
\cline{1-13}
 Period     & \multicolumn{3}{c|}{2017Q3 - 2020} & \multicolumn{3}{c|}{2021 - 2022} & \multicolumn{3}{c|}{2023} & \multicolumn{3}{c|}{2017Q3 - 2019Q2} \\
\cline{1-13}
 Coverage     & 50 \%       & 80 \%        & 98 \%        & 50\%        & 80\%        & 98 \%        & 50 \%     & 80 \%      & 98 \%      & 50  \%       & 80   \%     & 98 \%        \\
      \hline
HS    & 93.65      & 99.26      & 100.00     & 90.34     & 98.97     & 100.00     & 95.88   & 98.90    & 100.00   & 88.50      & 96.85     & 100.00     \\
CP    & 94.19      & 99.26      & 100.00     & 91.58     & 98.42     & 100.00     & 95.60   & 99.18    & 100.00   & 89.01      & 97.62     & 100.00     \\
QRA   & 95.52      & 99.18      & 100.00     & 94.04     & 98.84     & 100.00     & 96.70   & 99.73    & 100.00   & 92.87      & 98.14     & 99.87      \\
QRM   & 95.21      & 99.03      & 99.88      & 93.56     & 98.56     & 100.00     & 96.70   & 99.18    & 99.73    & 92.93      & 98.78     & 100.00     \\
FQRA  & 94.89      & 99.26      & 100.00     & 93.01     & 98.42     & 100.00     & 98.08   & 100.00   & 100.00   & 94.99      & 98.97     & 100.00     \\
FQRM  & 95.13      & 99.26      & 99.92      & 93.29     & 98.63     & 100.00     & 96.15   & 100.14   & 100.00   & 93.70      & 98.20     & 100.00     \\
sFQRA & 94.82      & 99.57      & 100.00     & 94.18     & 99.66     & 100.00     & 96.15   & 99.73    & 100.00   & 93.19      & 98.84     & 100.00     \\
sFQRM & 95.13      & 99.77      & 100.00     & 93.90     & 99.45     & 100.00     & 96.15   & 100.00   & 100.00   & 92.54      & 98.84     & 100.00    \\
\hline
\end{tabular}
}
\end{table}

\section{Conclusions}
\label{sec:conclusion}

This paper {describes a novel application of a factor-QRA method. In this approach, a rich set of almost 700 point predictions stemming from a single model calibrated to windows of different sizes is used for constructing prediction intervals. In order to overcome the colinearity problem and to efficiently use all available information, the data set is interpreted as a panel and its dynamics are summarized by common factors. The factors are next included in a quantile regression and used to predict future price quantiles constituting prediction intervals.

To assess the accuracy of the presented methods, three datasets from two European markets are analyzed: EPEX DA, EPEX ID and TGE. Their performance is evaluated over an out-of-sample period of more than 2 years. The data for EPEX day-ahead market has been additionally extended to include not only the calm period of 2017-2019, but also the COVID-19 pandemic and the period of the Ukrainian war. The instability of the market condition allows for a comprehensive assessment of the described approaches and increases the robustness of the presented results.

The performance of the proposed averaging schemes is compared with several well-established methods: historical simulations, conformal predictions, QRA and QRM. The results indicate that the factor-based methods, in particular, sFQRA and sFQRA, provide predictions that are not only more accurate than the benchmarks but also statistically reliable. They have empirical coverage close to the nominal level and pass the Christoffersen test for conditional coverage for the majority of hours and markets. We also compare two specifications of factor models that depend on data standardization. In contrast to other PCA applications, we normalize the data in the cross-sectional dimension rather than in the time dimension. This allows us to take into account both the level and the variability of the predictions across different window sizes. The results indicate that standardization has a huge impact on performance and leads to a significant improvement in the accuracy of the forecast.

Finally, forecasting methods are evaluated using economic measures. We propose an experiment that resembles a decision problem of an energy storage utility. We demonstrate how probabilistic forecasts can be used to build a trading strategy.
The results show that the data-driven strategies allow to decrease the trading volume in periods of low profitability and increase the average income per 1~MWh of trade. Since a lower number of charging cycles generates smaller variable costs, the proposed trading strategies, when properly used, dominate the benchmark unlimited bids approach.

The outcomes show additionally that an excellent statistical performance does not always correspond with the highest economic value of forecasts. In fact, we demonstrate that depending on the market and the price volatility regime, different top-performing models are revealed. 
While sFQRA/sFQRM and QRA/QRM approaches were placed among the best performing models for TGE dataset and low- and mid- volatility price regimes for the EPEX market (2017Q3-2020 and 2023), they underperform in the period with the most volatile prices (2021-2022). 
When considering the forecasts from the CP and HS benchmarks, they underperform all other approaches for the Polish market and rank among the worst models for the 2023 period for the German market. However, they perform well in the 2017Q3 - 2022 period for the EPEX dataset. 
These results emphasize the need of complementing the statistical evaluation of forecasts with economic measures that resemble real-life problems.

Combining quantile regression with dimension reduction techniques, such as PCA, allows large panels of point forecasts to be analyzed and used for probabilistic forecasting. We believe that this research can be further automatized, as in \cite{uni:mac:22}, and applied to different commodities and/or electricity markets.

\section*{Author contributions}
Conceptualization: K.M.; Investigation: T.S. and B.U.; Software: T.S. and B.U.;
Supervision: K.M.; Validation, K.M., T.S. and B.U.; Writing - original draft: T.S. and B.U.; Writing - review \& editing: K.M.

\section*{Acknowledgments}
This work was supported by the Ministry of Education and Science (MEiN, Poland) through Diamond Grant No. 0009/DIA/2020/49 (to T.S.) and the National Science Center (NCN, Poland) through MAESTRO grant No. 2018/30/A/HS4/00444 (to B.U.) and SONATA BIS grant No. 2019/34/E/HS4/00060 (to K.M).
 
\bibliography{lasso.bib}

\begin{thebibliography}{62}
\expandafter\ifx\csname natexlab\endcsname\relax\def\natexlab#1{#1}\fi
\providecommand{\url}[1]{\texttt{#1}}
\providecommand{\href}[2]{#2}
\providecommand{\path}[1]{#1}
\providecommand{\DOIprefix}{doi:}
\providecommand{\ArXivprefix}{arXiv:}
\providecommand{\URLprefix}{URL: }
\providecommand{\Pubmedprefix}{pmid:}
\providecommand{\doi}[1]{\href{http://dx.doi.org/#1}{\path{#1}}}
\providecommand{\Pubmed}[1]{\href{pmid:#1}{\path{#1}}}
\providecommand{\bibinfo}[2]{#2}
\ifx\xfnm\relax \def\xfnm[#1]{\unskip,\space#1}\fi
\bibitem[{Bai and Ng(2002)}]{bai:ng:02}
\bibinfo{author}{Bai, J.}, \bibinfo{author}{Ng, S.}, \bibinfo{year}{2002}.
\newblock \bibinfo{title}{Determining the number of factors in approximate
  factor models}.
\newblock \bibinfo{journal}{Econometrica} \bibinfo{volume}{70},
  \bibinfo{pages}{191--221}.
\bibitem[{Bill{\'e} et~al.(2023)Bill{\'e}, Gianfreda, Del~Grosso and
  Ravazzolo}]{bil:etal:23}
\bibinfo{author}{Bill{\'e}, A.G.}, \bibinfo{author}{Gianfreda, A.},
  \bibinfo{author}{Del~Grosso, F.}, \bibinfo{author}{Ravazzolo, F.},
  \bibinfo{year}{2023}.
\newblock \bibinfo{title}{Forecasting electricity prices with expert, linear,
  and nonlinear models}.
\newblock \bibinfo{journal}{International Journal of Forecasting}
  \bibinfo{volume}{39}, \bibinfo{pages}{570--586}.
\bibitem[{Bunn et~al.(2016)Bunn, Andresen, Chen and
  Westgaard}]{bun:and:che:wes:16}
\bibinfo{author}{Bunn, D.}, \bibinfo{author}{Andresen, A.},
  \bibinfo{author}{Chen, D.}, \bibinfo{author}{Westgaard, S.},
  \bibinfo{year}{2016}.
\newblock \bibinfo{title}{Analysis and forecasting of electricity price risks
  with quantile factor models}.
\newblock \bibinfo{journal}{Energy Journal} \bibinfo{volume}{37},
  \bibinfo{pages}{101--122}.
\bibitem[{Bunn et~al.(2018)Bunn, Gianfreda and Kermer}]{bun:gia:kre:18}
\bibinfo{author}{Bunn, D.}, \bibinfo{author}{Gianfreda, A.},
  \bibinfo{author}{Kermer, S.}, \bibinfo{year}{2018}.
\newblock \bibinfo{title}{A trading-based evaluation of density forecasts in a
  real-time electricity market}.
\newblock \bibinfo{journal}{Energies} \bibinfo{volume}{11},
  \bibinfo{pages}{2658}.
\bibitem[{Chatfield(1993)}]{cha:93}
\bibinfo{author}{Chatfield, C.}, \bibinfo{year}{1993}.
\newblock \bibinfo{title}{Calculating interval forecasts}.
\newblock \bibinfo{journal}{Journal of Business \& Economic Statistics}
  \bibinfo{volume}{11}, \bibinfo{pages}{121--135}.
\bibitem[{Christoffersen(1998)}]{chr:98}
\bibinfo{author}{Christoffersen, P.}, \bibinfo{year}{1998}.
\newblock \bibinfo{title}{Evaluating interval forecasts}.
\newblock \bibinfo{journal}{International Economic Review}
  \bibinfo{volume}{39}, \bibinfo{pages}{841--862}.
\bibitem[{Ciarreta et~al.(2022)Ciarreta, Muniain and Zarraga}]{cia:mun:zar:22}
\bibinfo{author}{Ciarreta, A.}, \bibinfo{author}{Muniain, P.},
  \bibinfo{author}{Zarraga, A.}, \bibinfo{year}{2022}.
\newblock \bibinfo{title}{Do jumps and cojumps matter for electricity price
  forecasting? evidence from the german-austrian day-ahead market}.
\newblock \bibinfo{journal}{Electric Power Systems Research}
  \bibinfo{volume}{212}, \bibinfo{pages}{108144}.
\bibitem[{Doostmohammadi et~al.(2017)Doostmohammadi, Amjady and
  Zareipour}]{doo:amj:zar:17}
\bibinfo{author}{Doostmohammadi, A.}, \bibinfo{author}{Amjady, N.},
  \bibinfo{author}{Zareipour, H.}, \bibinfo{year}{2017}.
\newblock \bibinfo{title}{Day-ahead financial loss/gain modeling and prediction
  for a generation company}.
\newblock \bibinfo{journal}{IEEE Transactions on Power Systems}
  \bibinfo{volume}{32}, \bibinfo{pages}{3360--3372}.
\bibitem[{Gammerman et~al.(1998)Gammerman, Vovk and Vapnik}]{vovk:gam:vap:98}
\bibinfo{author}{Gammerman, A.}, \bibinfo{author}{Vovk, V.},
  \bibinfo{author}{Vapnik, V.}, \bibinfo{year}{1998}.
\newblock \bibinfo{title}{Learning by transduction}, in:
  \bibinfo{booktitle}{Proceedings of the fourteenth conference on uncertainty
  in artificial intelligence}, \bibinfo{publisher}{Morgan Kaufmann Publishers
  Inc., San Francisco, CA}. p. \bibinfo{pages}{148–155}.
\bibitem[{Guo et~al.(2022)Guo, He, Liang and Ma}]{guo:22}
\bibinfo{author}{Guo, Y.}, \bibinfo{author}{He, F.}, \bibinfo{author}{Liang,
  C.}, \bibinfo{author}{Ma, F.}, \bibinfo{year}{2022}.
\newblock \bibinfo{title}{Oil price volatility predictability: New evidence
  from a scaled pca approach}.
\newblock \bibinfo{journal}{Energy Economics} \bibinfo{volume}{105},
  \bibinfo{pages}{105714}.
\bibitem[{He et~al.(2021)He, Zhang, Wen and Wang}]{he:21}
\bibinfo{author}{He, M.}, \bibinfo{author}{Zhang, Y.}, \bibinfo{author}{Wen,
  D.}, \bibinfo{author}{Wang, Y.}, \bibinfo{year}{2021}.
\newblock \bibinfo{title}{Forecasting crude oil prices: A scaled pca approach}.
\newblock \bibinfo{journal}{Energy Economics} \bibinfo{volume}{97},
  \bibinfo{pages}{105189}.
\bibitem[{Hong et~al.(2020)Hong, Pinson, Wang, Weron, Yang and
  Zareipour}]{hon:etal:20:OAJPE}
\bibinfo{author}{Hong, T.}, \bibinfo{author}{Pinson, P.},
  \bibinfo{author}{Wang, Y.}, \bibinfo{author}{Weron, R.},
  \bibinfo{author}{Yang, D.}, \bibinfo{author}{Zareipour, H.},
  \bibinfo{year}{2020}.
\newblock \bibinfo{title}{Energy forecasting: {A} review and outlook}.
\newblock \bibinfo{journal}{IEEE Open Access Journal of Power and Energy}
  \bibinfo{volume}{7}, \bibinfo{pages}{376--388}.
\bibitem[{Hubicka et~al.(2019)Hubicka, Marcjasz and Weron}]{hub:mar:wer:19}
\bibinfo{author}{Hubicka, K.}, \bibinfo{author}{Marcjasz, G.},
  \bibinfo{author}{Weron, R.}, \bibinfo{year}{2019}.
\newblock \bibinfo{title}{A note on averaging day-ahead electricity price
  forecasts across calibration windows}.
\newblock \bibinfo{journal}{IEEE Transactions on Sustainable Energy}
  \bibinfo{volume}{10}, \bibinfo{pages}{321--323}.
\bibitem[{Janczura et~al.(2013)Janczura, Tr\"uck, Weron and
  Wolff}]{jan:tru:wer:wol:13}
\bibinfo{author}{Janczura, J.}, \bibinfo{author}{Tr\"uck, S.},
  \bibinfo{author}{Weron, R.}, \bibinfo{author}{Wolff, R.},
  \bibinfo{year}{2013}.
\newblock \bibinfo{title}{Identifying spikes and seasonal components in
  electricity spot price data: A guide to robust modeling}.
\newblock \bibinfo{journal}{Energy Economics} \bibinfo{volume}{38},
  \bibinfo{pages}{96--110}.
\bibitem[{Janczura and W{\'o}jcik(2022)}]{jan:woj:22}
\bibinfo{author}{Janczura, J.}, \bibinfo{author}{W{\'o}jcik, E.},
  \bibinfo{year}{2022}.
\newblock \bibinfo{title}{Dynamic short-term risk management strategies for the
  choice of electricity market based on probabilistic forecasts of profit and
  risk measures. the german and the polish market case study}.
\newblock \bibinfo{journal}{Energy Economics} \bibinfo{volume}{110},
  \bibinfo{pages}{106015}.
\bibitem[{Jonsson et~al.(2013)Jonsson, Pinson, Nielsen, Madsen and
  Nielsen}]{jon:etal:13}
\bibinfo{author}{Jonsson, T.}, \bibinfo{author}{Pinson, P.},
  \bibinfo{author}{Nielsen, H.A.}, \bibinfo{author}{Madsen, H.},
  \bibinfo{author}{Nielsen, T.}, \bibinfo{year}{2013}.
\newblock \bibinfo{title}{Forecasting electricity spot prices accounting for
  wind power predictions}.
\newblock \bibinfo{journal}{IEEE Transactions on Sustainable Energy}
  \bibinfo{volume}{4}, \bibinfo{pages}{210--218}.
\bibitem[{Kath and Ziel(2018)}]{kat:zie:18}
\bibinfo{author}{Kath, C.}, \bibinfo{author}{Ziel, F.}, \bibinfo{year}{2018}.
\newblock \bibinfo{title}{The value of forecasts: Quantifying the economic
  gains of accurate quarter-hourly electricity price forecasts}.
\newblock \bibinfo{journal}{Energy Economics} \bibinfo{volume}{76},
  \bibinfo{pages}{411--423}.
\bibitem[{Kath and Ziel(2021)}]{kat:zie:21}
\bibinfo{author}{Kath, C.}, \bibinfo{author}{Ziel, F.}, \bibinfo{year}{2021}.
\newblock \bibinfo{title}{Conformal prediction interval estimation and
  applications to day-ahead and intraday power markets}.
\newblock \bibinfo{journal}{International Journal of Forecasting}
  \bibinfo{volume}{37}, \bibinfo{pages}{777--799}.
\bibitem[{Ketterer(2014)}]{ket:14}
\bibinfo{author}{Ketterer, J.}, \bibinfo{year}{2014}.
\newblock \bibinfo{title}{The impact of wind power generation on the
  electricity price in germany}.
\newblock \bibinfo{journal}{Energy Economics} \bibinfo{volume}{44},
  \bibinfo{pages}{270--280}.
\bibitem[{Koenker and Hallock(2001)}]{koe:hal:01}
\bibinfo{author}{Koenker, R.}, \bibinfo{author}{Hallock, K.F.},
  \bibinfo{year}{2001}.
\newblock \bibinfo{title}{Quantile regression}.
\newblock \bibinfo{journal}{Journal of Economic Perspectives}
  \bibinfo{volume}{15}, \bibinfo{pages}{143--156}.
\bibitem[{Koenker(2005)}]{koe:05}
\bibinfo{author}{Koenker, R.W.}, \bibinfo{year}{2005}.
\newblock \bibinfo{title}{Quantile Regression}.
\newblock \bibinfo{publisher}{Cambridge University Press}.
\bibitem[{Kolassa(2020)}]{kol:20}
\bibinfo{author}{Kolassa, S.}, \bibinfo{year}{2020}.
\newblock \bibinfo{title}{Why the “best” point forecast depends on the
  error or accuracy measure}.
\newblock \bibinfo{journal}{International Journal of Forecasting}
  \bibinfo{volume}{36}, \bibinfo{pages}{208--211}.
\bibitem[{Kupiec(1995)}]{kup:95}
\bibinfo{author}{Kupiec, P.H.}, \bibinfo{year}{1995}.
\newblock \bibinfo{title}{Techniques for verifying the accuracy of risk
  measurement models}.
\newblock \bibinfo{journal}{The Journal of Derivatives} \bibinfo{volume}{3},
  \bibinfo{pages}{73--84}.
\bibitem[{Lei et~al.(2018)Lei, G’Sell, Rinaldo, Tibshirani and
  Wasserman}]{lei:etal:18}
\bibinfo{author}{Lei, J.}, \bibinfo{author}{G’Sell, M.},
  \bibinfo{author}{Rinaldo, A.}, \bibinfo{author}{Tibshirani, R.J.},
  \bibinfo{author}{Wasserman, L.}, \bibinfo{year}{2018}.
\newblock \bibinfo{title}{Distribution-free predictive inference for
  regression}.
\newblock \bibinfo{journal}{Journal of the American Statistical Association}
  \bibinfo{volume}{113}, \bibinfo{pages}{1094--1111}.
\bibitem[{Li et~al.(2017)Li, Hurn and Clements}]{li:hu:cl:17}
\bibinfo{author}{Li, Z.}, \bibinfo{author}{Hurn, A.},
  \bibinfo{author}{Clements, A.}, \bibinfo{year}{2017}.
\newblock \bibinfo{title}{Forecasting quantiles of day-ahead electricity load}.
\newblock \bibinfo{journal}{Energy Economics} \bibinfo{volume}{67},
  \bibinfo{pages}{60--71}.
\bibitem[{Maciejowska(2020)}]{mac:20}
\bibinfo{author}{Maciejowska, K.}, \bibinfo{year}{2020}.
\newblock \bibinfo{title}{Assessing the impact of renewable energy sources on
  the electricity price level and variability -- a quantile regression
  approach}.
\newblock \bibinfo{journal}{Energy Economics} \bibinfo{volume}{85},
  \bibinfo{pages}{104532}.
\bibitem[{Maciejowska et~al.(2021)Maciejowska, Nitka and
  Weron}]{mac:nit:wer:21}
\bibinfo{author}{Maciejowska, K.}, \bibinfo{author}{Nitka, W.},
  \bibinfo{author}{Weron, T.}, \bibinfo{year}{2021}.
\newblock \bibinfo{title}{Enhancing load, wind and solar generation for
  day-ahead forecasting of electricity prices}.
\newblock \bibinfo{journal}{Energy Economics} \bibinfo{volume}{99},
  \bibinfo{pages}{105273}.
\bibitem[{Maciejowska et~al.(2016)Maciejowska, Nowotarski and
  Weron}]{mac:now:wer:16}
\bibinfo{author}{Maciejowska, K.}, \bibinfo{author}{Nowotarski, J.},
  \bibinfo{author}{Weron, R.}, \bibinfo{year}{2016}.
\newblock \bibinfo{title}{Probabilistic forecasting of electricity spot prices
  using {F}actor {Q}uantile {R}egression {A}veraging}.
\newblock \bibinfo{journal}{International Journal of Forecasting}
  \bibinfo{volume}{32}, \bibinfo{pages}{957--965}.
\bibitem[{Maciejowska et~al.(2020)Maciejowska, Uniejewski and
  Serafin}]{mac:uni:ser:20}
\bibinfo{author}{Maciejowska, K.}, \bibinfo{author}{Uniejewski, B.},
  \bibinfo{author}{Serafin, T.}, \bibinfo{year}{2020}.
\newblock \bibinfo{title}{Pca forecast averaging—predicting day-ahead and
  intraday electricity prices}.
\newblock \bibinfo{journal}{Energies} \bibinfo{volume}{13},
  \bibinfo{pages}{3530}.
\bibitem[{Mansouri et~al.(2023)Mansouri, Abolmasoumi and Ghadimi}]{man:etal:23}
\bibinfo{author}{Mansouri, A.}, \bibinfo{author}{Abolmasoumi, A.H.},
  \bibinfo{author}{Ghadimi, A.A.}, \bibinfo{year}{2023}.
\newblock \bibinfo{title}{Weather sensitive short term load forecasting using
  dynamic mode decomposition with control}.
\newblock \bibinfo{journal}{Electric Power Systems Research}
  \bibinfo{volume}{221}, \bibinfo{pages}{109387}.
\bibitem[{Marcjasz et~al.(2023)Marcjasz, Narajewski, Weron and
  Ziel}]{mar:nar:wer:zie:23}
\bibinfo{author}{Marcjasz, G.}, \bibinfo{author}{Narajewski, M.},
  \bibinfo{author}{Weron, R.}, \bibinfo{author}{Ziel, F.},
  \bibinfo{year}{2023}.
\newblock \bibinfo{title}{Distributional neural networks for electricity price
  forecasting}.
\newblock \bibinfo{journal}{Energy Economics} \bibinfo{volume}{125},
  \bibinfo{pages}{106843}.
\bibitem[{Marcjasz et~al.(2018)Marcjasz, Serafin and Weron}]{mar:ser:wer:18}
\bibinfo{author}{Marcjasz, G.}, \bibinfo{author}{Serafin, T.},
  \bibinfo{author}{Weron, R.}, \bibinfo{year}{2018}.
\newblock \bibinfo{title}{Selection of calibration windows for day-ahead
  electricity price forecasting}.
\newblock \bibinfo{journal}{Energies} \bibinfo{volume}{11},
  \bibinfo{pages}{2364}.
\bibitem[{Marcjasz et~al.(2019)Marcjasz, Uniejewski and
  Weron}]{mar:uni:wer:19:narx}
\bibinfo{author}{Marcjasz, G.}, \bibinfo{author}{Uniejewski, B.},
  \bibinfo{author}{Weron, R.}, \bibinfo{year}{2019}.
\newblock \bibinfo{title}{On the importance of the long-term seasonal component
  in day-ahead electricity price forecasting with {NARX} neural networks}.
\newblock \bibinfo{journal}{International Journal of Forecasting}
  \bibinfo{volume}{35}, \bibinfo{pages}{1520--1532}.
\bibitem[{Marcjasz et~al.(2020)Marcjasz, Uniejewski and Weron}]{mar:uni:wer:20}
\bibinfo{author}{Marcjasz, G.}, \bibinfo{author}{Uniejewski, B.},
  \bibinfo{author}{Weron, R.}, \bibinfo{year}{2020}.
\newblock \bibinfo{title}{Probabilistic electricity price forecasting with
  {NARX} networks: Combine point or probabilistic forecasts?}
\newblock \bibinfo{journal}{International Journal of Forecasting}
  \bibinfo{volume}{36}, \bibinfo{pages}{466--479}.
\bibitem[{Mashlakov et~al.(2021)Mashlakov, Kuronen, Lensu, Kaarna and
  Honkapuro}]{mas:etal:21}
\bibinfo{author}{Mashlakov, A.}, \bibinfo{author}{Kuronen, T.},
  \bibinfo{author}{Lensu, L.}, \bibinfo{author}{Kaarna, A.},
  \bibinfo{author}{Honkapuro, S.}, \bibinfo{year}{2021}.
\newblock \bibinfo{title}{Assessing the performance of deep learning models for
  multivariate probabilistic energy forecasting}.
\newblock \bibinfo{journal}{Applied Energy} \bibinfo{volume}{285},
  \bibinfo{pages}{116405}.
\bibitem[{Narajewski and Ziel(2020)}]{nar:zie:20JCM}
\bibinfo{author}{Narajewski, M.}, \bibinfo{author}{Ziel, F.},
  \bibinfo{year}{2020}.
\newblock \bibinfo{title}{Econometric modelling and forecasting of intraday
  electricity prices}.
\newblock \bibinfo{journal}{Journal of Commodity Markets} \bibinfo{volume}{19},
  \bibinfo{pages}{100107}.
\bibitem[{Narajewski and Ziel(2022)}]{nar:ze:22}
\bibinfo{author}{Narajewski, M.}, \bibinfo{author}{Ziel, F.},
  \bibinfo{year}{2022}.
\newblock \bibinfo{title}{Optimal bidding in hourly and quarter-hourly
  electricity price auctions: Trading large volumes of power with market impact
  and transaction costs}.
\newblock \bibinfo{journal}{Energy Economics} \bibinfo{volume}{110},
  \bibinfo{pages}{105974}.
\bibitem[{Nowotarski and Weron(2015)}]{now:wer:15}
\bibinfo{author}{Nowotarski, J.}, \bibinfo{author}{Weron, R.},
  \bibinfo{year}{2015}.
\newblock \bibinfo{title}{Computing electricity spot price prediction intervals
  using quantile regression and forecast averaging}.
\newblock \bibinfo{journal}{Computational Statistics} \bibinfo{volume}{30},
  \bibinfo{pages}{791--803}.
\bibitem[{Nowotarski and Weron(2018)}]{now:wer:18}
\bibinfo{author}{Nowotarski, J.}, \bibinfo{author}{Weron, R.},
  \bibinfo{year}{2018}.
\newblock \bibinfo{title}{Recent advances in electricity price forecasting: {A}
  review of probabilistic forecasting}.
\newblock \bibinfo{journal}{Renewable and Sustainable Energy Reviews}
  \bibinfo{volume}{81}, \bibinfo{pages}{1548--1568}.
\bibitem[{Petropoulos et~al.(2022)Petropoulos, Apiletti, Assimakopoulos and
  \textit{et al.}}]{pet:etal:22}
\bibinfo{author}{Petropoulos, F.}, \bibinfo{author}{Apiletti, D.},
  \bibinfo{author}{Assimakopoulos, V.}, \bibinfo{author}{\textit{et al.}},
  \bibinfo{year}{2022}.
\newblock \bibinfo{title}{Forecasting: theory and practice}.
\newblock \bibinfo{journal}{International Journal of Forecasting}
  \bibinfo{volume}{38}, \bibinfo{pages}{705--871}.
\bibitem[{Serafin et~al.(2022)Serafin, Marcjasz and Weron}]{ser:mar:wer:22}
\bibinfo{author}{Serafin, T.}, \bibinfo{author}{Marcjasz, G.},
  \bibinfo{author}{Weron, R.}, \bibinfo{year}{2022}.
\newblock \bibinfo{title}{Trading on short-term path forecasts of intraday
  electricity prices}.
\newblock \bibinfo{journal}{Energy Economics} \bibinfo{volume}{112},
  \bibinfo{pages}{106125}.
\bibitem[{Serafin et~al.(2019)Serafin, Uniejewski and Weron}]{ser:uni:wer:19}
\bibinfo{author}{Serafin, T.}, \bibinfo{author}{Uniejewski, B.},
  \bibinfo{author}{Weron, R.}, \bibinfo{year}{2019}.
\newblock \bibinfo{title}{Averaging predictive distributions across calibration
  windows for day-ahead electricity price forecasting}.
\newblock \bibinfo{journal}{Energies} \bibinfo{volume}{12},
  \bibinfo{pages}{256}.
\bibitem[{Stock and Watson(2002)}]{sto:wat:02a}
\bibinfo{author}{Stock, J.H.}, \bibinfo{author}{Watson, M.W.},
  \bibinfo{year}{2002}.
\newblock \bibinfo{title}{Forecasting using principal components from a large
  number of predictors}.
\newblock \bibinfo{journal}{Journal of the American Statistical Association}
  \bibinfo{volume}{97}, \bibinfo{pages}{1167--1179}.
\bibitem[{Uniejewski(2024)}]{uni:22}
\bibinfo{author}{Uniejewski, B.}, \bibinfo{year}{2024}.
\newblock \bibinfo{title}{Electricity price forecasting with smoothing quantile
  regression averaging: Quantifying economic benefits of probabilistic
  forecasts} \bibinfo{note}{DOI arXiv: 10.48550/arXiv.2302.00411}.
\bibitem[{Uniejewski and Maciejowska(2022)}]{uni:mac:22}
\bibinfo{author}{Uniejewski, B.}, \bibinfo{author}{Maciejowska, K.},
  \bibinfo{year}{2022}.
\newblock \bibinfo{title}{Lasso principal component averaging: A fully
  automated approach for point forecast pooling}.
\newblock \bibinfo{journal}{International Journal of Forecasting} .
\bibitem[{Uniejewski et~al.(2016)Uniejewski, Nowotarski and
  Weron}]{uni:now:wer:16}
\bibinfo{author}{Uniejewski, B.}, \bibinfo{author}{Nowotarski, J.},
  \bibinfo{author}{Weron, R.}, \bibinfo{year}{2016}.
\newblock \bibinfo{title}{Automated variable selection and shrinkage for
  day-ahead electricity price forecasting}.
\newblock \bibinfo{journal}{Energies} \bibinfo{volume}{9},
  \bibinfo{pages}{621}.
\bibitem[{Uniejewski and Weron(2021)}]{uni:wer:21}
\bibinfo{author}{Uniejewski, B.}, \bibinfo{author}{Weron, R.},
  \bibinfo{year}{2021}.
\newblock \bibinfo{title}{Regularized quantile regression averaging for
  probabilistic electricity price forecasting}.
\newblock \bibinfo{journal}{Energy Economics} \bibinfo{volume}{95},
  \bibinfo{pages}{105121}.
\bibitem[{Uniejewski et~al.(2018)Uniejewski, Weron and Ziel}]{uni:wer:zie:18}
\bibinfo{author}{Uniejewski, B.}, \bibinfo{author}{Weron, R.},
  \bibinfo{author}{Ziel, F.}, \bibinfo{year}{2018}.
\newblock \bibinfo{title}{Variance stabilizing transformations for electricity
  spot price forecasting}.
\newblock \bibinfo{journal}{IEEE Transactions on Power Systems}
  \bibinfo{volume}{33}, \bibinfo{pages}{2219--2229}.
\bibitem[{Velliangiri et~al.(2019)Velliangiri, Alagumuthukrishnan and
  {Thankumar joseph}}]{vel:19}
\bibinfo{author}{Velliangiri, S.}, \bibinfo{author}{Alagumuthukrishnan, S.},
  \bibinfo{author}{{Thankumar joseph}, S.I.}, \bibinfo{year}{2019}.
\newblock \bibinfo{title}{A review of dimensionality reduction techniques for
  efficient computation}.
\newblock \bibinfo{journal}{Procedia Computer Science} \bibinfo{volume}{165},
  \bibinfo{pages}{104--111}.
\bibitem[{Vovk et~al.(2009)Vovk, Nouretdinov and Gammerman}]{vovk:etal:09}
\bibinfo{author}{Vovk, V.}, \bibinfo{author}{Nouretdinov, I.},
  \bibinfo{author}{Gammerman, A.}, \bibinfo{year}{2009}.
\newblock \bibinfo{title}{{On-line predictive linear regression}}.
\newblock \bibinfo{journal}{The Annals of Statistics} \bibinfo{volume}{37},
  \bibinfo{pages}{1566 -- 1590}.
\bibitem[{Vovk et~al.(2018)Vovk, Nouretdinov, Manokhin and
  Gammerman}]{vovk:etal:18}
\bibinfo{author}{Vovk, V.}, \bibinfo{author}{Nouretdinov, I.},
  \bibinfo{author}{Manokhin, V.}, \bibinfo{author}{Gammerman, A.},
  \bibinfo{year}{2018}.
\newblock \bibinfo{title}{Cross-conformal predictive distributions}, in:
  \bibinfo{editor}{Gammerman, A.}, \bibinfo{editor}{Vovk, V.},
  \bibinfo{editor}{Luo, Z.}, \bibinfo{editor}{Smirnov, E.},
  \bibinfo{editor}{Peeters, R.} (Eds.), \bibinfo{booktitle}{Proceedings of the
  Seventh Workshop on Conformal and Probabilistic Prediction and Applications},
  \bibinfo{publisher}{PMLR}. pp. \bibinfo{pages}{37--51}.
\bibitem[{Vovk et~al.(2019)Vovk, Shen and Manokhin}]{vovk:etal:19}
\bibinfo{author}{Vovk, V.}, \bibinfo{author}{Shen, J.},
  \bibinfo{author}{Manokhin, V.}, \bibinfo{year}{2019}.
\newblock \bibinfo{title}{Nonparametric predictive distributions based on
  conformal prediction}.
\newblock \bibinfo{journal}{Machine Learning} \bibinfo{volume}{108},
  \bibinfo{pages}{445–474}.
\bibitem[{Wan et~al.(2014a)Wan, Xu, Pinson, Dong and Wong}]{wan:etal:14c}
\bibinfo{author}{Wan, C.}, \bibinfo{author}{Xu, Z.}, \bibinfo{author}{Pinson,
  P.}, \bibinfo{author}{Dong, Z.}, \bibinfo{author}{Wong, K.},
  \bibinfo{year}{2014}a.
\newblock \bibinfo{title}{Probabilistic forecasting of wind power generation
  using extreme learning machine}.
\newblock \bibinfo{journal}{IEEE Transactions on Power Systems}
  \bibinfo{volume}{29}, \bibinfo{pages}{1033--1044}.
\bibitem[{Wan et~al.(2014b)Wan, Xu, Wang, Dong and Wong}]{wan:etal:14b}
\bibinfo{author}{Wan, C.}, \bibinfo{author}{Xu, Z.}, \bibinfo{author}{Wang,
  Y.}, \bibinfo{author}{Dong, Z.}, \bibinfo{author}{Wong, K.},
  \bibinfo{year}{2014}b.
\newblock \bibinfo{title}{A hybrid approach for probabilistic forecasting of
  electricity price}.
\newblock \bibinfo{journal}{IEEE Transactions on Smart Grid}
  \bibinfo{volume}{5}, \bibinfo{pages}{463--470}.
\bibitem[{Wang et~al.(2023)Wang, Wang, Wang, Wang, He and Zhang}]{wan:etal:23}
\bibinfo{author}{Wang, D.}, \bibinfo{author}{Wang, P.}, \bibinfo{author}{Wang,
  P.}, \bibinfo{author}{Wang, C.}, \bibinfo{author}{He, Z.},
  \bibinfo{author}{Zhang, W.}, \bibinfo{year}{2023}.
\newblock \bibinfo{title}{Probabilistic prediction with locally weighted
  jackknife predictive system}.
\newblock \bibinfo{journal}{Complex \& Intelligent Systems}
  \bibinfo{volume}{9}, \bibinfo{pages}{5761--5778}.
\bibitem[{Wang et~al.(2019)Wang, Zhang, Tan, Hong, Kirschen and
  Kang}]{wan:etal:19}
\bibinfo{author}{Wang, Y.}, \bibinfo{author}{Zhang, N.}, \bibinfo{author}{Tan,
  Y.}, \bibinfo{author}{Hong, T.}, \bibinfo{author}{Kirschen, D.},
  \bibinfo{author}{Kang, C.}, \bibinfo{year}{2019}.
\newblock \bibinfo{title}{Combining probabilistic load forecasts}.
\newblock \bibinfo{journal}{IEEE Transactions on Smart Grid}
  \bibinfo{volume}{10}, \bibinfo{pages}{3664--3674}.
\bibitem[{Weron(2006)}]{wer:06}
\bibinfo{author}{Weron, R.}, \bibinfo{year}{2006}.
\newblock \bibinfo{title}{Modeling and Forecasting Electricity Loads and
  Prices: A Statistical Approach}.
\newblock \bibinfo{publisher}{John Wiley {\&} Sons, Chichester}.
\bibitem[{Weron(2014)}]{wer:14}
\bibinfo{author}{Weron, R.}, \bibinfo{year}{2014}.
\newblock \bibinfo{title}{Electricity price forecasting: {A} review of the
  state-of-the-art with a look into the future}.
\newblock \bibinfo{journal}{International Journal of Forecasting}
  \bibinfo{volume}{30}, \bibinfo{pages}{1030--1081}.
\bibitem[{Weron and Ziel(2018)}]{wer:zie:18}
\bibinfo{author}{Weron, R.}, \bibinfo{author}{Ziel, F.}, \bibinfo{year}{2018}.
\newblock \bibinfo{title}{Forecasting Electricity Prices: A Guide to Robust
  Modeling}.
\newblock \bibinfo{publisher}{CRC Press}.
\newblock \bibinfo{note}{Forthcoming}.
\bibitem[{Zareipour et~al.(2010)Zareipour, Canizares and
  Bhattacharya}]{zar:can:bha:10}
\bibinfo{author}{Zareipour, H.}, \bibinfo{author}{Canizares, C.A.},
  \bibinfo{author}{Bhattacharya, K.}, \bibinfo{year}{2010}.
\newblock \bibinfo{title}{Economic impact of electricity market price
  forecasting errors: A demand-side analysis}.
\newblock \bibinfo{journal}{IEEE Transactions on Power Systems}
  \bibinfo{volume}{25}, \bibinfo{pages}{254--262}.
\bibitem[{Zhang et~al.(2022)Zhang, Tang, Wu, Du and Chen}]{zha:etal:22}
\bibinfo{author}{Zhang, T.}, \bibinfo{author}{Tang, Z.}, \bibinfo{author}{Wu,
  J.}, \bibinfo{author}{Du, X.}, \bibinfo{author}{Chen, K.},
  \bibinfo{year}{2022}.
\newblock \bibinfo{title}{Short term electricity price forecasting using a new
  hybrid model based on two-layer decomposition technique and ensemble
  learning}.
\newblock \bibinfo{journal}{Electric Power Systems Research}
  \bibinfo{volume}{205}, \bibinfo{pages}{107762}.
\bibitem[{Zhang et~al.(2018)Zhang, Quan and Srinivasan}]{zha:qua:sri:18}
\bibinfo{author}{Zhang, W.}, \bibinfo{author}{Quan, H.},
  \bibinfo{author}{Srinivasan, D.}, \bibinfo{year}{2018}.
\newblock \bibinfo{title}{Parallel and reliable probabilistic load forecasting
  via quantile regression forest and quantile determination}.
\newblock \bibinfo{journal}{Energy} \bibinfo{volume}{160},
  \bibinfo{pages}{810--819}.

\end{thebibliography}

\end{document}